\documentclass[sigconf, nonacm]{acmart}

\usepackage{balance}

\usepackage{url}
\usepackage[labelfont=bf,textfont=bf]{caption}
\usepackage{subcaption}
\usepackage{graphicx}
\usepackage{xcolor}
\usepackage{xspace}
\usepackage{array}
\usepackage[noend]{algpseudocode}
\usepackage{dblfloatfix}
\usepackage{balance}
\usepackage{algorithm}
\usepackage{mathtools}
\usepackage{amsmath}
\usepackage[nodisplayskipstretch]{setspace}

\newcommand{\longonly}[1]{\empty{#1}}
\newcommand{\shortonly}[1]{\empty{}}

\newcommand{\reffigure}[1]{Figure~\ref{#1}}
\newcommand{\refsection}[1]{Section~\ref{#1}}

\newcommand{\reftable}[1]{Table~\ref{#1}}
\newcommand{\refequation}[1]{Equation~\ref{#1}}
\newcommand{\refappendix}[1]{Appendix}

\newcommand{\zap}[1]{}
\newcommand{\btree}{B$^+$-tree\xspace}
\newcommand{\btrees}{B$^+$-trees\xspace}

\newcommand{\revise}[1]{{#1}}

\newcommand{\minus}{\scalebox{0.75}[1.0]{$-$}}

\newcommand{\btreeshort}{B$^+$\xspace}

\newcommand{\static}{\revise{\emph{\btreeshort-static}}\xspace}
\newcommand{\statictuned}{\emph{\btreeshort-static-tuned}\xspace}
\newcommand{\dynamic}{\revise{\emph{\btreeshort-dynamic}}\xspace}
\newcommand{\dynamicmem}{\revise{\emph{\btreeshort-dynamic-MEM}}\xspace}

\newcommand{\accordionindex}{\emph{Accordion-index}\xspace}
\newcommand{\accordiondata}{\emph{Accordion-data} \xspace}

\newcommand{\partition}{\emph{Partitioned}\xspace}
\newcommand{\partitionmem}{\emph{Partitioned-MEM}\xspace}

\newcommand{\partitionopt}{\emph{Partitioned-OPT}\xspace}

\newcommand\vldbdoi{XX.XX/XXX.XX}
\newcommand\vldbpages{XXX-XXX}
\newcommand\vldbvolume{14}
\newcommand\vldbissue{1}
\newcommand\vldbyear{2020}
\newcommand\vldbauthors{\authors}
\newcommand\vldbtitle{\shorttitle} 

\begin{document}
	
\shortonly{
\title{Breaking Down Memory Walls: Adaptive Memory Management in LSM-based Storage Systems}
}
\longonly{
\title{Breaking Down Memory Walls: Adaptive Memory Management in LSM-based Storage Systems (Extended Version)}
}

\author{Chen Luo}
\affiliation{%
	\institution{University of California, Irvine}
}
\email{cluo8@uci.edu}

\author{Michael J. Carey}
\affiliation{%
	\institution{University of California, Irvine}
}
\email{mjcarey@ics.uci.edu}

\begin{abstract}
Log-Structured Merge-trees (LSM-trees) have been widely used in modern NoSQL systems.
Due to their out-of-place update design, LSM-trees have introduced memory walls 
among the memory components of multiple LSM-trees and between the write memory and the buffer cache.
Optimal memory allocation among these regions is non-trivial because it is highly workload-dependent.
Existing LSM-tree implementations instead adopt static memory allocation schemes due to their simplicity and robustness, sacrificing performance.
In this paper, we attempt to break down these memory walls in LSM-based storage systems.
We first present a memory management architecture that enables adaptive memory management.
We then present a partitioned memory component structure with new flush policies to better exploit the write memory to minimize the write cost.
To break down the memory wall between the write memory and the buffer cache,
we further introduce a memory tuner that tunes the memory allocation between these two regions.
We have conducted extensive experiments in the context of Apache AsterixDB using the YCSB and TPC-C benchmarks and we present the results here.
\end{abstract}

\maketitle

\shortonly{
\begingroup\small\noindent\raggedright\textbf{PVLDB Reference Format:}\\
\vldbauthors. \vldbtitle. PVLDB, \vldbvolume(\vldbissue): \vldbpages, \vldbyear.\\
\href{https://doi.org/\vldbdoi}{doi:\vldbdoi}
\endgroup
\begingroup
\renewcommand\thefootnote{}\footnote{\noindent
	This work is licensed under the Creative Commons BY-NC-ND 4.0 International License. Visit \url{https://creativecommons.org/licenses/by-nc-nd/4.0/} to view a copy of this license. For any use beyond those covered by this license, obtain permission by emailing \href{mailto:info@vldb.org}{info@vldb.org}. Copyright is held by the owner/author(s). Publication rights licensed to the VLDB Endowment. \\
	\raggedright Proceedings of the VLDB Endowment, Vol. \vldbvolume, No. \vldbissue\ %
	ISSN 2150-8097. \\
	\href{https://doi.org/\vldbdoi}{doi:\vldbdoi} \\
}\addtocounter{footnote}{-1}\endgroup
}

\section{Introduction}
Log-Structured Merge-trees (LSM-trees)~\cite{lsm1996} are widely used in modern NoSQL systems,
such as LevelDB~\cite{leveldb}, RocksDB~\cite{rocksdb}, Cassandra~\cite{cassandra}, HBase~\cite{hbase},
X-Engine~\cite{xengine}, and AsterixDB~\cite{asterixdb-web}.
Unlike traditional in-place update structures, LSM-trees adopt an out-of-place update design by first buffering all writes in memory;
they are subsequently flushed to disk to form immutable disk components.
The disk components are periodically merged to improve query performance and reclaim space occupied by obsolete records.

Efficient memory management is critical for storage systems to achieve optimal performance.
Compared to update-in-place systems where all pages are managed within shared buffer pools,
LSM-trees have introduced additional memory walls\footnote{
\revise{In this paper, the term \emph{memory wall} refers to the barriers among various memory regions that prevent efficient memory sharing.}}.
Due to the LSM-tree's out-of-place update nature, the write memory is isolated from the buffer cache.
\revise{Moreover, data management systems adopting LSM storage engines, such as MyRocks~\cite{myrocks} on RocksDB~\cite{rocksdb},
	PolarDB~\cite{polardb2018} on X-Engine~\cite{xengine}, and AsterixDB~\cite{asterixdb2014}, must deal with
	multiple heterogeneous LSM-trees from multiple datasets and indexes.
This requires the write memory to be efficiently shared among multiple LSM-trees.}
Since the optimal memory allocation heavily depends on the workload,
memory management should be workload-adaptive to maximize the system performance.

Unfortunately, adaptivity is non-trivial, as it is highly workload-dependent.
Existing LSM-tree implementations, such as RocksDB~\cite{rocksdb} and AsterixDB~\cite{asterixdb-memory},
have opted for simplicity and robustness over optimal performance by adopting static memory allocation schemes.
For example, RocksDB sets a static size limit (default 64MB) for each memory component.
AsterixDB specifies the maximum number N of writable datasets (default 8) so that each active dataset,
including its primary and secondary indexes, receives 1/N of the total write memory.
Both systems allocate separate static budgets for the write memory and the buffer cache.
\revise{Despite their simplicity and robustness, static memory allocation schemes may negatively impact the system performance and efficiency
	due to sub-optimal memory allocation.}

\revise{\textbf{Our Contributions.}}
In this paper, we seek to break down these memory walls in LSM-based storage systems to \revise{enable adaptive memory memory management and maximize performance and efficiency}.
\revise{As the first contribution, we present an adaptive memory management architecture for LSM-based storage systems.} In this architecture, the overall memory budget is divided into the write memory region and the buffer cache region.
Within the write memory region, the memory allocation of each memory component is purely driven by its demands, i.e., write rates,
to minimize the overall write amplification.
The two regions are connected via a \emph{memory tuner} that adaptively tunes the memory allocation between the write memory and the buffer cache.

\revise{
	As the second contribution, we present a series of techniques for efficiently managing the write memory for LSM-trees in order to minimize their write I/O cost. We first present a new LSM memory component structure for managing the write memory for a single LSM-tree, and then propose
	novel flush policies for managing the write memory for multiple LSM-trees.
}

Our third contribution is the detailed design and implementation of a memory tuner that adaptively
tunes the memory allocation between the write memory and the buffer cache to reduce the system's overall I/O cost.
The memory tuner performs on-line tuning by modeling the I/O cost of LSM-trees without any a priori knowledge of the workload.
This further allows the memory tuner to quickly adjust the memory allocation when the workload changes.

\revise{As the last contribution,} we have implemented all of the proposed techniques inside Apache AsterixDB~\cite{asterixdb-web}.
We have carried out extensive experiments on both the YCSB benchmark~\cite{ycsb2010} and the TPC-C benchmark~\cite{tpcc}
to evaluate the effectiveness of the proposed techniques.
The experimental results show that the proposed techniques successfully reduce the disk I/O cost, which in turn maximizes system efficiency and overall performance.

The remainder of this paper is organized as follows.
\refsection{sec:background} discusses background information and related work.
\refsection{sec:architecture} presents our adaptive memory management architecture for LSM-trees.
\refsection{sec:write-memory} describes the new memory component structure for managing the write memory.
\refsection{sec:memory-tuner} presents the design and implementation of the memory tuner.
\refsection{sec:experiment} experimentally evaluates the proposed techniques.
Finally, \refsection{sec:conclusion} concludes the paper.

\section{Background}
\label{sec:background}

\subsection{Log-Structured Merge Trees}
\label{sec:background-lsm}
The LSM-tree~\cite{lsm1996} is a persistent index structure optimized for write-intensive workloads.
LSM-trees perform out-of-place updates by always buffering writes into a memory component and appending log records to a transaction log for durability.
Writes are flushed to disk when either the memory component is full, called a \emph{memory-triggered flush},
or when the transaction log length becomes too long, called a \emph{log-triggered flush}.

A query over an LSM-tree has to reconcile the entries with identical keys from multiple components, as entries from newer components override those from older components.
A range query searches all components simultaneously using a priority queue to perform reconciliation.
A point lookup query simply searches all components from newest to oldest until the first match is found.
To speed up point lookups, a common optimization is to build Bloom filters~\cite{bloom-filter1970} over the sets of keys stored in disk components.

To improve query performance and space utilization, disk components are periodically merged according a pre-defined merge policy. In practice, two types of merge policies are commonly used~\cite{lsm-survey}, both of which organize disk components into ``levels''.
The leveling merge policy maintains one component per level.
When a component at Level $i$ is $T$ times larger than that of Level $i-1$, it will be merged into Level $i+1$ to form a new component.
In contrast, the tiering merge policy maintains $T$ components per level. When a Level $i$ becomes full with $T$ components, they are merged together into a new component at Level $i+1$.

\begin{figure}
	\centering
	\includegraphics[width=\linewidth]{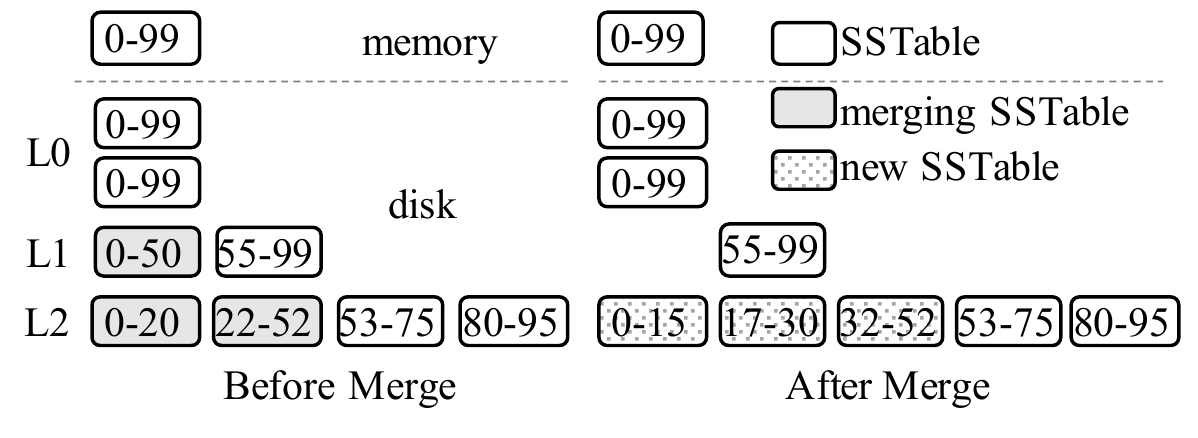}
	\vspace{-0.3in}
	\caption{Example Partitioned LSM-tree}
	\label{fig:partitioned-lsm}
\end{figure}

\textbf{Partitioning.} In practice, a common optimization is to range-partition a disk component into multiple (often fixed-size) SSTables\footnote{\revise{An SSTable (Sorted String Table)~\cite{bigtable} stores a set of immutable rows sorted on keys.}} to bound the processing time and temporary space of each merge.
This optimization is often used together with the leveling merge policy, as pioneered by LevelDB~\cite{leveldb}.
An example of a partitioned LSM-tree with the leveling merge policy is shown in \reffigure{fig:partitioned-lsm},
where each SSTable is labeled with its key range.
Note that $L_0$ is not partitioned since its SSTables are directly flushed from memory.
$L_0$ also stores multiple SSTables with overlapping key ranges to absorb write bursts.
To merge an SSTable from $L_i$ to $L_{i+1}$, all of its overlapping SSTables at $L_{i+1}$ are selected and
these SSTables are merged to form new SSTables at $L_{i+1}$.
For example in \reffigure{fig:partitioned-lsm}, the SSTable labeled 0-50 at $L_1$ will be merged
with the SSTables labeled 0-20 and 22-52 at $L_2$,
which produce new SSTables labeled 0-15, 17-30, and 32-52 at $L_2$.
When the LSM-tree becomes too large, a new level must be added.
To maximize space utilization, the new level should be added at $L_1$ instead of the last level~\cite{rocksdb-space2017}.
The last level is always treated as full, which in turn determines the maximum sizes of other levels.
When the maximum size of $L_1$ is larger than $T$ times the write memory size (or the configured base level size), a new $L_1$ is added
while all remaining levels $L_i$ become $L_{i+1}$.
In our work, we will focus on the partitioned leveling structure due to its wide adoption in today's LSM-based systems.

\begin{table}
	\caption{LSM-tree Notation}
	\label{table:lsm-notation}
	\centering
	{
		\begin{tabular}{|c|m{12.5em}|m{5.5em}|}
			\hline
			\textbf{Notation} & \textbf{Definition} & \textbf{Example} \\
			\hline	 \multicolumn{3}{|c|}{Global Notation} \\
			\hline		$T$  & size ratio of the merge policy & 10 \\
			\hline		$P$ & disk page size & 4 KB/page \\
			\hline		$M_{w}$ & total write memory size & 1GB \\
			\hline	 \multicolumn{3}{|c|}{Local Notation} \\
			\hline		$e_i$ & entry size & 100 B/entry \\
			\hline		$a_i$ & ratio of an LSM-tree's write memory to total write memory & 20\% \\
			\hline		$N_i$ & number of levels (excluding $L_0$)& 3 \\
			\hline		$|L_{l_i}| $ & size of  Level $L_{l_i}$ & 10 GB \\
			\hline		$C_i$ & write I/O cost per entry & 4 pages/entry\\
			\hline
		\end{tabular}
	}
\end{table}

\textbf{Write Memory vs. Write Cost.}
Here we provide a simple cost analysis to show the relationship between the write memory size and the per-entry write I/O cost.
Our notation is shown in \reftable{table:lsm-notation}.
Note that since we consider multiple LSM-trees, \reftable{table:lsm-notation} contains global notation that is valid for all LSM-trees
and local notation that is specific to one LSM-tree.
In the remainder of this paper, for the $i$-th LSM-tree,
we add the subscript $i$ to denote the local notation for this LSM-tree.
Note that we have further introduced the notation $a$ to denote the write memory ratio of an LSM-tree.
Thus, for the $i$-th LSM-tree, its write memory size is $a_i\cdot M_w$.
Moreover, given a collection of $K$ LSM-trees, we have $\sum_{i=1}^{K}a_i = 1$.

Each entry written to an LSM-tree is flushed to disk once and merged multiple times down to the last level.
The per-entry flush cost is $\frac{e}{P}$ pages/entry.
Merging an SSTable at Level $L_i$ usually has $T$ overlapping SSTables at Level $L_{i+1}$.
Thus, to merge an entry from $L_0$ to the last level, the overall merge cost is $\frac{e}{P}\cdot (T+1) \cdot N$ pages/entry.
Here the number of levels $N$ can be expressed using other terms as follows.
Given an LSM-tree whose write memory size is $a \cdot M_w$, the maximum size of $i$-th level is $a\cdot M_w \cdot T^i$.
Based on the size of the last Level $|L_N|$, we have $|L_N| \le a\cdot M_w \cdot T^N$.
Thus, $N$ can be approximated as $\log_{T}{\frac{|L_N|}{a\cdot M_w}}$.
Putting everything together, the per-entry write cost $C$ is approximately
\begin{equation}
\label{eq:write-cost}
C = \frac{e}{P} + \frac{e}{P}\cdot (T+1) \cdot \log_{T}{\frac{|L_N|}{a\cdot M_w}}
\end{equation}
As \refequation{eq:write-cost} shows, a larger write memory reduces the write cost by reducing the number of disk levels.
Thus, it is important to utilize a large write memory efficiently to reduce the write cost.

\subsection{Apache AsterixDB}
\label{sec:asterixdb}
Apache AsterixDB~\cite{asterixdb-web, asterixdb2014,asterixdb2019} is a parallel, semi-structured Big Data Management System (BDMS)
for efficiently managing large amounts of data.
It supports a feed-based framework for efficient data ingestion~\cite{asterixdb-feed2015,idea2019}.
The records of a dataset in AsterixDB are hash-partitioned based on their primary keys across multiple nodes of a shared-nothing cluster~\cite{asterixdb-storage2014}.
\revise{Each partition of a dataset uses a primary LSM-tree to store the data records with each component begin organized as a \btree.}
Local secondary indexes, including LSM-based \btrees, R-trees, and inverted indexes, can also be built to expedite query processing.

AsterixDB uses a static memory allocation scheme for simplicity and robustness~\cite{asterixdb-memory}.
It specifies static memory budgets for the buffer cache and the write memory.
Moreover, AsterixDB specifies the maximum number D of writable datasets (default 8) so that each active dataset receives 1/D of the total write memory.
When a dataset's write memory is full, all of its LSM-trees, including its primary index and secondary indexes, will be flushed to disk together.
If the user writes to the D+1-st dataset, the least recently written active dataset will be evicted to reclaim its write memory.
In this work, we use AsterixDB as a testbed to evaluate the proposed techniques and compare them to other baselines.

\subsection{Related Work}
\label{sec:related-work}
\textbf{LSM-trees.}
Recently, a large number of improvements have been proposed to optimize the original~\cite{lsm1996} LSM-tree design.
These improvements include optimizing write performance~\cite{triad2017,hailstorm2020,dostoevsky2018,lsm-bush,hashkv2019,wisckey2017,lsm-append2019,sifrdb2018,pebblesdb2017,fpga-lsm2020},
supporting auto-tuning of LSM-trees~\cite{monkey2017,monkey-tods,lsm-model2016,lethe2020},
optimizing LSM-based secondary indexes and filters~\cite{lsm-storage2019,secondary2018,rosetta2020},
minimizing write stalls~\cite{silk2019, lsm-stability2019, blsm2012},
and extending the applicability of LSM-trees~\cite{umzi2019,slimdb2017}.
We refer readers to a recent survey~\cite{lsm-survey} for a more detailed description of these LSM-tree improvements.

In terms of memory management, FloDB~\cite{flodb2017} presents a two-level memory component structure to mask write latencies.
However, it mainly optimizes for peak in-memory throughput instead of reducing the overall write cost. 
Accordion~\cite{accordion2018} introduces a multi-level memory component structure with memory flushes and merges.
One drawback is that Accordion does not range-partition memory components, resulting in high memory utilization during large memory merges.
We will further experimentally evaluate Accordion in \refsection{sec:experiment}. 
Monkey~\cite{monkey2017} uses analytical models to tune the memory allocation between memory components and Bloom filters.
ElasticBF~\cite{elasticbf2019} proposes a dynamic Bloom filter management scheme to adjust false positives rates based on the data hotness.
Different from Monkey and ElasticBF, in our work Bloom filters are managed the same paged way as SSTables through the buffer cache.
It should also be noted that virtually all previous research only considers the memory management of a single LSM-tree.

\textbf{Database Memory Management.}
The importance of memory management, or buffer management, has long been recognized for database systems.
Various buffer replacement policies, such as DBMIN~\cite{dbmin1986}, 2Q~\cite{2q}, LRU-K~\cite{lruk1993}, and Hot-Set~\cite{hot-set1982},
have been proposed to reduce buffer cache misses.
These replacement policies are orthogonal to this work because we mainly focus on the memory walls introduced by the LSM-tree's out-of-place update design.

Automatic memory tuning is also an important problem for database systems.
Some commercial DBMSs have supported auto-tuning the memory allocation
among different memory regions~\cite{sqlserver2005,db2-memory2006}.
Depending on the tuning goals, the memory tuning techniques can be classified as maximizing the overall throughput or meeting latency requirements.
DB2's self-tuning memory manager (STMM)~\cite{db2-memory2006} is an example of the former, using control theory to tune the memory allocation.
However, STMM targets targets a traditional in-place update system,
which does not include the write memory used by LSM-trees.
For the latter, the relationship between the buffer cache size and the cache miss rate must be predicted,
using either analytical models~\cite{tune-formula2008} or machine learning approaches~\cite{ibtune2019}.

There has been recent interest in exploiting machine learning to tune database configurations~\cite{ituned2009,tuning2019,tune-ml2017,tune-rl2019},
where memory allocation is treated as one tuning knob.
These approaches usually require additional training steps and user inputs.
Different from these approaches, our memory tuner uses a white-box approach; it carefully models the I/O cost of LSM-based storage systems.

\section{Memory Management Architecture}
\label{sec:architecture}
\begin{figure}
	\centering
	\includegraphics[width=\linewidth]{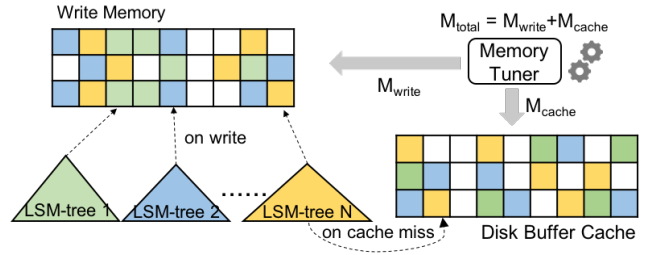}
	\vspace{-0.25in}
	\caption{Memory Management Architecture}
	\label{fig:memory-architecture}
\end{figure}

In this section, we present our memory management architecture to enable adaptive memory management.
In this architecture, depicted in \reffigure{fig:memory-architecture},
the total memory budget is divided into the write memory $M_{write}$ and the buffer cache $M_{cache}$.
These two regions are further connected via a memory tuner, which periodically performs memory tuning to reduce the total I/O cost.

\textbf{Write Memory.} 
The write memory stores incoming writes for all LSM-trees.
To maximize memory utilization, we do not set static size limits for the individual memory components.
Instead, all memory components are managed through a shared memory pool.
When an LSM-tree has insufficient memory to store its incoming writes, more pages will be requested from the pool.
When the overall write memory usage is too high, an LSM-tree is selected to have its memory component flushed to disk.

While the basic idea of this design is straightforward, there are several technical challenges here.
First, how can we best utilize the write memory to minimize the write cost?
Second, since the memory component of an LSM-tree now becomes dynamic,
how can we adjust the disk levels as the write memory changes to always make optimal performance trade-offs?
Finally, given a collection of heterogeneous LSM-trees of different sizes, how can we allocate the write memory to these LSM-trees to minimize the overall write cost?
We will present our solutions to these challenges in \refsection{sec:write-memory}.

\textbf{Buffer Cache.}
\revise{The buffer cache stores (immutable) disk pages of the SSTables as well as their Bloom filters for all LSM-trees.
Even though all LSM-trees share the same buffer cache, their merges are performed separately.}
As in traditional database systems, all disk pages are managed together using a predefined buffer replacement policy.
For example, AsterixDB uses the clock replacement policy to manage its shared buffer cache.
In this work, we mainly focus on the memory allocation given to the buffer cache instead of cache replacement within the buffer cache.

\textbf{Memory Tuner.}
Given a memory budget, the memory tuner attempts to tune the memory allocation between
the write memory and the buffer cache to reduce the total I/O cost.
The key property of the memory tuner is that it takes a white-box approach by carefully modeling the I/O cost of LSM-based storage systems and thus does not require any offline training.
We will describe the design and implementation of the memory tuner in \refsection{sec:memory-tuner}.

\section{Managing Write Memory}
\label{sec:write-memory}
Now we present our solution for managing the write memory.
We first describe the memory component structure of a single LSM-tree and then present techniques for managing multiple LSM-trees.

\subsection{Partitioned Memory Component}
\label{sec:partitioned-memory-component}
\begin{figure}
	\centering
	\includegraphics[width=\linewidth]{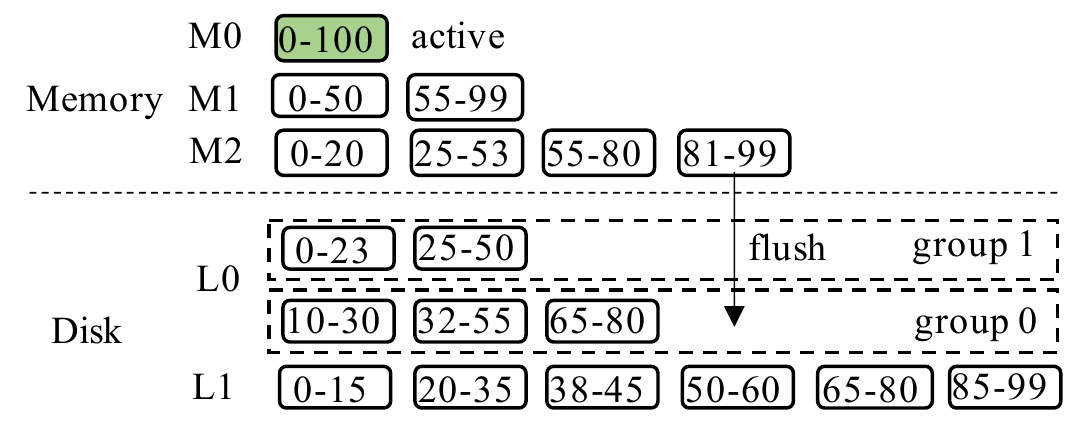}
	\vspace{-0.25in}
	\caption{LSM-tree with a Partitioned Memory Component}
	\label{fig:partitioned-memory}
\end{figure}

\revise{
\subsubsection{Basic Design}
With the new memory management architecture, a memory component can become very large since its size is not limited.
Existing LSM-tree implementations use skiplists or \btrees to manage memory components and always flush a memory component entirely to disk.
However, this reduces memory utilization for two reasons.
First, an updatable \btree has internal fragmentation, as its pages are about 2/3 full~\cite{btree-space-1978}.
Second, after a flush a large chunk of write memory will be freed (vacated) all at once, which reduces the average memory utilization over time.
\footnote{\revise{To see this, consider a single LSM-tree with a large memory component. If its memory component is flushed entirely, the average memory utilization over time will be less than 50\%.}}

To maximize the memory utilization, we propose to use a partitioned in-memory LSM-tree to manage the write memory,
which is called a \emph{partitioned memory component}.
An example LSM-tree with this structure is shown in \reffigure{fig:partitioned-memory},
which has an active SSTable at $M_0$ for storing incoming writes and a set of memory levels for storing immutable SSTables.
When a memory level $M_i$ is full, one of its SSTables is merged into the next level $M_{i+1}$ using a \emph{memory merge}.
We use a greedy selection policy to select SSTables to merge by minimizing the ratio between the size of the overlapping SSTables at $M_{i+1}$ and the size of the selected SSTable at $M_i$, i.e., the overlapping ratio.
Memory SSTables must be eventually flushed to disk.
For a memory-triggered flush, SSTables at the last memory level ($M_2$ in \reffigure{fig:partitioned-memory}) are flushed to disk in a round-robin way.
For a log-triggered flush, the SSTable with the minimum log sequence number (LSN) will be flushed (as well as all overlapping SSTables at higher levels) to facilitate log truncation.

Compared to the monolithic memory component structure used in existing systems, the proposed structure increases the memory utilization and reduces the write amplification in several ways.
First, an LSM-tree achieves much higher space utilization than \btrees.
For example, with a size ratio of 10, an LSM-tree achieves 90\% space utilization, which is much higher than that of a \btree.
Moreover, since the proposed structure is range-partitioned, it can naturally flush one memory SSTable at a time using partial flushes so that the write memory stays full.
Finally, partial flushes further reduce the write amplification by creating skews at the last level~\cite{lsm-model2016}.
Since SSTables are flushed in a round-robin way, the flushed SSTable will have received the most updates.
Thus, the key ranges of these SSTables will be denser than the average, which reduces the subsequent merge cost.
In the remainder of this section, we further discuss the detailed design of the proposed structure.
}

\revise{
\subsubsection{Grouped $L_0$}
In the original LSM-tree design (\reffigure{fig:partitioned-lsm}), the disk level $L_0$ stores a list of (unpartitioned) SSTables ordered by their recency.
When the number of SSTables at $L_0$ exceeds a pre-defined threshold, flushes will be paused to bound the worse-case query performance.
Multiple $L_0$ SSTables are also merged together into $L_1$ to reduce the merge cost.
In the partitioned memory component structure, where the flushed SSTables are range partitioned,
the original $L_0$ structure is unsuitable since non-overlapping SSTables have no negative impact on queries.
Thus, flushes should only be paused when there are too many overlapping SSTables.

To better accommodate the new memory component structure, we propose a new $L_0$ structure by organizing its SSTables into groups,
where each group contains a set of disjoint SSTables.}
Groups are ordered based on their recency, where the keys in a newer group override the keys in an older group.
When the total number of groups at $L_0$ exceeds a predefined threshold, incoming flushes will be stopped.
\revise{We further use the following heuristics to reduce the number of groups at $L_0$ and also the write amplification.}
First, when an SSTable is flushed to disk, it is always inserted into the oldest possible group where all newer groups do not have any overlapping SSTables.
Otherwise, if no such group can be found, a new group is created.
\longonly{Consider the two groups in \reffigure{fig:partitioned-memory}, where group 0 is older than group 1.
When flushing the SSTable labeled 81-99, the resulting SSTable will be inserted into the older group 0.
If the SSTable labeled 25-53 is flushed, a new group will be created because group 1 contains an overlapping SSTable 25-50.}
Second, SSTables from the smallest group that contains the fewest SSTables will be merged into $L_1$ first.
Specifically, an SSTable from this group as well as any overlapping SSTables from other $L_0$ groups are merged with the overlapping SSTables at $L_1$.
To reduce write amplification, the merging SSTable is selected to minimize the ratio between the total size of the overlapping SSTables at $L_1$ and the total size of the merging SSTables at $L_0$.
\longonly{Consider the example LSM-tree in \reffigure{fig:partitioned-memory}.
Group 1 will be selected for the merge because it has fewer SSTables than group 0.
The SSTable labeled 0-23 could be merged with the SSTable labeled 10-30 at group 0 and the SSTables labeled 0-15 and 20-35 at $L_1$, whose overlapping ratio would be $1$.
The SSTable labeled 25-50 could be merged with the SSTables labeled 10-30 and 32-55 in group 0 and the SSTables labeled
from 0-15 to 50-60 at $L_1$, whose overlapping ratio would be $4/3$.
Thus, to reduce write amplification, the SSTable labeled 0-23 will be selected for the merge.
}

\zap{
One potential issue with the above design is that memory merges and flushes may lead to deadlocks.
To see this problem, consider an extreme situation where all SSTables at the last memory level are merging and the total write memory is full.
As a result, memory merges cannot proceed until some write memory is reclaimed by flushes.
However, flushes also cannot proceed because all of the last-level SSTables are merging.
Since deadlocks are rare, our solution is to break deadlocks when they occur.
Specifically, when the write memory is full and no flush can proceed, some memory merges that are blocking flushes will be aborted.
}

\subsubsection{Adjusting Disk Levels}
In the new memory component architecture, the write memory of each LSM-tree is allocated on-demand and is thus dynamic.
Since the write cost of an LSM-tree depends on the number of disk levels, the number of disk levels needs to be adjusted as its write memory size changes\footnote{Our preliminary solution~\cite{lsm-memory2020} was to only increase the number of disk on-levels without ever decreasing it.
	However, our subsequent evaluation showed that this led to 5\%-10\% performance loss compared to an optimal LSM-tree.}.

Recall that to maximize the space utilization, levels are only added or deleted at $L_1$.
For each disk level $L_i$, its maximum size is $a \cdot M_w \cdot T^i$.
Here we assume that the size of each disk level $|L_i|$ is relatively stable,
but the write memory allocated to an LSM-tree $a \cdot M_w$ is dynamic.
When an LSM-tree's write memory size becomes too small, i.e., $a\cdot M_w \cdot T < |L_1|$, a new $L_1$ should be added to reduce the write cost.
In this case, an empty $L_1$ can be added and all remaining levels $L_i$ simply become $L_{i+1}$.
In contrast, when the write memory size becomes too big, i.e., $a\cdot M_w \cdot T > |L_2|$,
$L_1$ becomes redundant and can be deleted. 
However, implementing this strategy directly can cause oscillation when the write memory is close to this threshold.
To avoid this, the deletion of $L_1$ can be delayed until the write memory further grows by a factor of $f$, i.e., $a \cdot M_w \cdot T > f \cdot |L_2|$.
As we will see in \refsection{sec:experiment}, delaying the deletion of $L_1$ has a much smaller impact than delaying the addition of a level.
In general, a larger $f$ better avoids oscillation but may have a larger negative impact on write amplification.
By default, we set $f$ to $1.5$ to balance these two factors.

\begin{figure}
	\centering
	\includegraphics[width=0.85\linewidth]{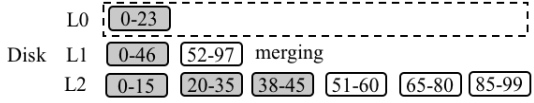}
	\vspace{-0.1in}
	\caption{Example Merge for Removing $L_1$}
	\label{fig:remove-level}
\end{figure}

To delete $L_1$, all existing SSTables from $L_1$ must be merged into $L_2$.
Here we describe an efficient solution to delete $L_1$ smoothly with minimal overhead.
To delete $L_1$, SSTables from $L_0$ can be directly merged into $L_2$ along with all overlapping SSTables at $L_1$.
Consider the example in \reffigure{fig:remove-level}.
To delete $L_1$, the SSTable labeled 0-23 at $L_0$ and the SSTable labeled 0-46 at $L_1$ can be directly merged into $L_2$.
This mechanism ensures that $L_1$ will not receive new SSTables but does not itself guarantee that $L_1$ will eventually become empty.
To address this problem, low-priority merges are also scheduled to merge SSTables from $L_1$ into $L_2$ when there are no merges at other levels.
These two operations ensure that $L_1$ will eventually become empty, and it can then be removed from the LSM-tree.

\subsubsection{Partial Flush vs. Full Flush}
As mentioned before, the new memory component structures enables partial flushes, i.e., flushing one SSTable at a time.
While possible, partial flushes may not always be an optimal choice.
Consider the case when the total write memory is large and flushes are only triggered by log truncation.
Since the oldest entries can be distributed across all memory SSTables, most memory SSTables may have to be flushed in order to truncate the log.
\revise{If a \emph{full flush} is performed, which will merge-sort all memory SSTables across all levels, the flushed SSTables will have non-overlapping key ranges.
In contrast, if partial flushes are used, the flushed SSTables may have overlapping key ranges, which will require subsequent merges to make these SSTables fully sorted
and thus incur extra merge I/O cost.}
Thus, for a log-triggered flush, the optimal flush choice depends on the write memory size and the maximum transaction log length.

Developing an optimal flush solution is non-trivial since it also heavily depends on the key distribution of the write workload.
Here we propose a simple heuristic to dynamically switch between partial and full flushes for log-triggered flushes.
The basic idea is to use a window to keep track of how much write memory has been partially flushed before the log-triggered flush, where the window size is set as the maximum transaction log length.
\revise{When log truncation is needed, if a large amount of write memory has already been flushed before, then only a small number of remaining SSTables will need to be flushed and thus partial flushes will be a better choice.
	Otherwise, full flushes should be performed.
Implementation-wise, we introduce a threshold parameter $\beta$ ranging from 0 to 1.
When the total amount of previously flushed write memory is larger than $\beta$ times the total write memory, partial flushes will be performed.
Otherwise, the entire memory component will be flushed together using a full flush.}
Based on some preliminary simulation results, we set our default value for $\beta$ to be $0.5$ to minimize the overall write cost.
(We leave the further exploration of the optimal choice of partial and full flushes as future work.)

\subsection{Managing Multiple LSM-trees}
\label{sec:multiple-lsm-tree}
When managing multiple LSM-trees, a fundamental question is how to allocate portions of the write memory to these LSM-trees.
Since write memory is allocated on-demand, this question becomes how to select LSM-trees to flush.
For log-triggered flushes, the LSM-tree with the minimum LSN should be flushed to perform log truncation.
For memory-triggered flushes, existing LSM-tree implementations, such as RocksDB~\cite{rocksdb} and HBase~\cite{hbase},
choose to flush the LSM-tree with the largest memory component.
We call this policy the \emph{max-memory} flush policy.
The intuition is that flushing this LSM-tree can reclaim the most write memory, which can be used for subsequent writes.
However, this policy may not be suitable for our partitioned memory components
because flushing any LSM-tree will reclaim the same amount of write memory due to partial SSTable flushes.

\textbf{Min-LSN Policy}.
One alternative flush policy is to always flush the LSM-tree with the minimum LSN for both log-triggered and memory-triggered flushes.
We call this policy the \emph{min-LSN} flush policy.
The intuition is that the flush rate of an LSM-tree should be approximately proportional to its write rate.
A hotter LSM-tree should be flushed more often than a colder one, but it still receives more write memory.
This policy also facilitates log truncation, which can be beneficial if flushes are dominated by log truncation.

\textbf{Optimal Policy}.
Given a collection of $K$ LSM-trees, our ultimate goal is to find an optimal memory allocation that minimizes the overall write cost.
For the $i$-th LSM-tree, we denote $r_i$ as its the write rate (bytes/s).
The optimal memory allocation can be obtained by solving the following optimization problem:
\begin{equation}
\label{eq:memory-opt}
\min_{a_i}\sum_{i=1}^{K}{\frac{r_i}{e_i}\cdot C_i}\textrm{, s.t.} \sum_{i=1}^{K}{a_i} = 1
\end{equation}
By substituting \refequation{eq:write-cost} from \refsection{sec:background-lsm} into \refequation{eq:memory-opt} and using the Lagrange multiplier method, 
the optimal write memory ratio $a^{opt}_i$ for the $i$-th LSM-tree is $a^{opt}_i = r_i/\sum_{j=1}^{K}{r_j}$.
This shows that the write memory allocated to each LSM-tree should be proportional to its write rate.
We call this policy the \emph{optimal} flush policy.
In terms of its implementation,
we can use a window to keep track of the total number of writes to each LSM-tree, where the window size is set as the maximum transaction log length.
When a memory-triggered flush is requested, each active LSM-tree is checked in turn and a flush is scheduled if
its write memory ratio $a_i$ is larger than its optimal write memory ratio $a^{opt}_i$.

\section{Memory Tuner}
\label{sec:memory-tuner}
After discussing how to efficiently manage the write memory, we now proceed to describe the memory tuner to tune the memory allocation between the write memory and the buffer cache.
We first provide an overview of the tuning approach, which is followed by its design and implementation.

\begin{figure}[b]
	\centering
	\includegraphics[width=0.75\linewidth]{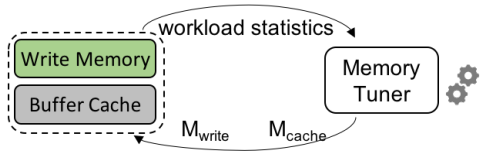}
	\vspace{-0.1in}
	\caption{Workflow of Memory Tuner}
	\label{fig:memory-tuner}
\end{figure}

\subsection{Tuning Approach}
The goal of the memory tuner is to find an optimal memory allocation between the write memory and the buffer cache to minimize the I/O cost per operation.
This should in turn maximize the system efficiency as well as the overall throughput.
Suppose the total available memory is $M$.
For ease of discussion, let us assume the write memory size is $x$, which implies that the buffer cache size is $M-x$.
Let $write(x)$ and $read(x)$ be the write cost and read cost per operation when the write memory is $x$.
Our tuning goal is to minimize the weighted I/O cost per operation (pages/op) $cost(x) = \omega\cdot write(x) + \gamma\cdot read(x)$.
The \revise{non-negative} weights $\omega$ and $\gamma$ allow users to adjust the tuning goal for different use cases.
For example, on hard disks, $\omega$ can be set smaller since LSM-trees mainly use sequential I/Os for writes,
while on SSDs $\omega$ can be set larger since SSD writes are often more expensive than SSD reads.

\revise{In order to minimize the tuning goal $cost(x)$, we use an online gradient descent approach to adaptively tune the memory allocation based on $cost'(x)$,
	which is $\omega\cdot write'(x) + \gamma\cdot read'(x)$.
Intuitively, $cost'(x)$ measures how the I/O cost changes if more write memory is allocated.
Based on $cost'(x)$, the memory tuner can tune the memory allocation accordingly to reduce $cost(x)$.
It should be noted that the optimality of the memory tuner depends on the shape of $cost(x)$.
We will discuss this issue further in \refsection{sec:tuner-optimality}.}

Based on this idea, the memory tuner uses a feedback-control loop to tune memory allocation, as depicted in \reffigure{fig:memory-tuner}.
The memory tuner continuously uses the collected statistics to \revise{tune} the memory allocation between the write memory and the buffer cache
without any user input nor training samples.
Before describing the details of the memory tuner, 
we first introduce some notation used by the memory tuner (\reftable{table:memory-tune-notation}) in addition to the LSM-tree notation listed in \reftable{table:lsm-notation}.
Note that with secondary indexes each operation may write multiple entries to multiple LSM-trees.

\begin{table}
	\caption{Memory Tuner Notation}
	\label{table:memory-tune-notation}
	\centering
	{
		\begin{tabular}{|c|m{13.5em}|m{4.5em}|}
			\hline
			\textbf{Notation} & \textbf{Definition} & \textbf{Example} \\
			\hline \multicolumn{3}{|c|}{Global Notation}\\
			\hline 		$K$ & number of LSM-trees & 8 \\
			\hline		$op$  & number of operations observed & 10K ops\\
			\hline 		$saved_q$ & saved query disk I/O by the simulated cache & 0.01 page/op\\
			\hline 		$saved_m$ & saved merge disk I/O by the simulated cache & 0.002 page/op \\
			\hline 		$sim$ & simulated cache size & 32 MB \\
			\hline \multicolumn{3}{|c|}{Local Notation} \\
			\hline 		$w_i$  & number of entries written to an LSM-tree & 50K entries \\
			\hline 		$flush_{log_i}$ & write memory flushed by log truncation & 1 GB \\
			\hline		$flush_{mem_i}$ &  write memory flushed by high memory usage & 8 GB \\
			\hline
		\end{tabular}
	}
\end{table}

\subsection{Estimating the Write Cost Derivative}
For the $i$-th LSM-tree, recall that \refequation{eq:write-cost} computes the per-entry write cost $C_i$.
Since each operation writes $\frac{w_i}{op}$ entries to this LSM-tree,
its write cost per operation $write_i(x)$ can computed as $\frac{w_i}{op}\cdot C_i$.
By taking the derivative of $write_i(x)$, we have
\begin{equation}
\label{eq:merge-cost-derivative-i}
write'_i(x)= \frac{w_i}{op}  \cdot \frac{e_i}{P}\cdot (T+1)\cdot \frac{1}{x\cdot \ln{T}}
\end{equation}
To reduce the estimation error, instead of collecting statistics for $op$, $w_i$, $e_i$ and $P$,
we simply collect the total number of merge writes per operation, $merge_i(x)$, in the last tuning cycle.
By substituting $merge_i(x)$ into \refequation{eq:merge-cost-derivative-i}, we have
\begin{equation}
\label{eq:merge-cost-derivative-i-simplify}
write'_i(x) = -\frac{merge_i(x)}{x \cdot \ln{\frac{|L_{N_i}|}{a_i\cdot x}}}
\end{equation}

Here we assume that the write memory of an LSM-tree is always smaller than its last level size.
Thus, the estimated value of $write'_i(x)$ in \refequation{eq:merge-cost-derivative-i-simplify} is always negative as long as $merge_i(x)$ is not zero.
This implies that adding more write memory can always reduce the write cost, which may not hold in practice.
Once flushes are dominated by log truncation, adding more write memory will not further reduce the write cost.
To account for the impact of log-triggered flushes, we further multiply \refequation{eq:merge-cost-derivative-i-simplify}
by a scale factor $\frac{flush_{mem_i}}{flush_{mem_i}+ flush_{log_i} }$ that we also keep statistics for.
Intuitively, this scale factor will be close to 1 if flushes are mainly triggered by high memory usage
and it will approach to 0 if flushes are mostly triggered by log truncation.
Finally, $write'(x)$ is the sum of $write'_i(x)$ for all LSM-trees:
\begin{equation}
\label{eq:merge-cost-all}
write'(x) = \sum_{i=1}^{K}{-\frac{merge_i(x)}{x \cdot \ln{\frac{|L_{N_i}|}{a_i\cdot x}}} \cdot \frac{flush_{mem_i}}{flush_{mem_i}+ flush_{log_i} } }  
\end{equation}

\longonly{
\textbf{Example 5.1.}
Consider an example with two LSM-trees.
Suppose that the total write memory $x$ is 128MB.
Suppose that the first LSM-tree receives 80\% of the write memory ($a_1 = 0.8$) with a last level size of 100GB ($|L_{N_1}| = 100GB$)
and that its merge cost per operation is 1 page/op ($merge_1(128MB) = 1$ page/op).
Similarly, for the second LSM-tree, suppose that $a_2=0.2$, $|L_{N_2}|=50GB$, and $merge_2(128MB) = 0.8$ page/op.
For simplicity, suppose that all flushes are memory-triggered.
Based on \refequation{eq:merge-cost-derivative-i-simplify}, $write'_1(128MB) \approx \minus1.08e^{\minus9}$ page/op and $write'_2(128MB) \approx  \minus0.78 e^{\minus9}$ page/op. Thus, $write'(128MB) \approx \minus1.86e^{\minus9}$ page/op.
This implies that if we allocate one more byte of write memory,
the write cost can be reduced by $1.86 e^{\minus9}$ page/op.
}

\subsection{Estimating the Read Cost Derivative}
\revise{$read'(x)$ measures how the read cost per operation changes if more write memory is allocated.
Since disk reads are performed by both queries and merges,}
we rewrite $read(x) = read_q(x) + read_m(x)$, where $read_q(x)$ is the number of query disk reads per operation
and $read_m(x)$ is the number of merge disk reads per operation.

\revise{$read'_q(x)$ measures the impact of larger write memory on the query read cost, as larger write memory increases the buffer cache miss rate.
To estimate $read'_q(x)$, we use a simulated cache as suggested by~\cite{db2-memory2006}.}
This simulated cache only stores page IDs.
Whenever a page is evicted from the buffer cache, its page ID is added to the simulated cache.
Whenever a page is about to be read from disk, a disk I/O could have been saved if the simulated cache contains that page ID.
Suppose that the simulated cache size is $sim$ and the saved read cost per operation is $saved_q$,
then $read'_q(x) = \frac{saved_q}{sim}$.

\revise{$read'_m(x)$ measures the impact of larger write memory on the merge read cost.
Intuitively, larger write memory reduces the disk merge cost, but also increases the buffer cache miss rate.}
Thus, to estimate $read'_m(x)$, we first rewrite $read_m(x) = pin_m(x) \cdot miss_m(x)$.
\revise{$pin_m(x)$ is the number of page pins \revise{performed by} disk merges per operation and it can be obtained by collecting runtime statistics.
$miss_m(x)$ is the cache miss rate for merges and it can be computed as $misss_m(x) = \frac{read_m(x)}{pin_m(x)}$.}
Based on the derivative rule, we have $read'_m(x) = pin'_m(x) \cdot miss_m(x) + pin_m(x) \cdot miss'_m(x)$.
$pin'_m(x)$ is the number of saved merge page pins per unit of write memory.
Recall that we have computed $write'(x)$, which is the number of saved disk writes per unit of write memory.
On average, each merge disk write requires $\frac{pin_m(x)}{merge(x)}$ page pins.
As a result, $pin'_m(x) = write'(x) \cdot \frac{pin_m(x)}{merge(x)}$.
To estimate $miss'_m(x)$, we again use the simulated cache to estimate the number of saved merge reads per operation $saved_m$.
Thus, $miss'_m(x) = \frac{saved_m}{pin_m(x) \cdot sim}$.
Putting everything together, $read'_m(x) =  write'(x) \cdot \frac{read_m(x)}{merge(x)} + \frac{saved_m}{sim}$.

Finally, $read'(x)$ can be computed as
\begin{equation}
\label{eq:read-cost}
read'(x) = \frac{saved_q + saved_m}{sim} + write'(x) \cdot \frac{read_m(x)}{merge_m(x)}
\end{equation}

\longonly{
\textbf{Example 5.2.}
	Continuing from Example 5.1, suppose that the simulated cache size is 32MB.
	Suppose that the simulated cache reports that the saved query disk reads per operation is $saved_q=0.01$ page/op
	and that the saved merge disk reads per operation is $saved_m=0.008$ page/op.
	Moreover, suppose that the total number of merge disk reads per operation is $read_m(x)=2.4$ page/op.
	Thus, we can compute $read'(x) = -1.94e^{\minus9}$ page/op.
	This means that allocating 1 more byte of write memory can decrease the disk read cost per operation by $1.94e^{\minus9}$ page/op,
	as disk reads are mainly performed by merges in this case.
}

\subsection{\revise{Tuning} Memory Allocation}
\label{sec:memory-tuner-finding}
\revise{Based on the computed $cost'(x)$, the memory allocation can then be tuned to reduce $cost(x)$.
Intuitively, the write memory size $x$ should be decreased if $cost'(x) > 0$ and it should be increased if $cost'(x) < 0$.
To speed up the tuning process, we use the Newton–Raphson method to find the root of $cost'(x)$ directly,
as $cost'(x) = 0$ is a necessary condition for minimizing $cost(x)$.
$cost'(x)$ is approximated as a linear function $cost'(x) = Ax + B$ using the last $K$ memory allocations, where $K$ by default is set to 3.
Given the current write memory size $x_i$, the next value is computed as $x_{i+1} = x_i - \frac{cost'(x_i)}{A}$.
}

\revise{Since the memory tuner deals with a complex system with constantly changing workloads and possible estimation errors,
several heuristics are used to ensure the stability of the memory tuner.
First, when the tuner does not have enough samples to construct the linear function or when the estimated memory allocation $x_{i+1}$ does not reduce the total cost, we simply fall back to a fixed step size, e.g., 5\% of the total memory.}
Second, the maximum step size is limited based on the memory region whose memory needs to be decreased.
The intuition is that taking memory from a region may be harmful because both the write memory and the buffer cache are subject to diminishing returns.
Thus, at each tuning step, we limit the maximum decreased memory size for either memory region to 10\% of its currently allocated memory size.
Finally, the memory tuner uses two stopping criteria to avoid oscillation.
The memory allocation is not changed if the step size is too small, e.g., smaller than 32MB, or if the expected cost reduction is too small, e.g., smaller than 0.1\% of the current I/O cost.

The last question for implementing the memory tuner is determining the appropriate tuning cycle length.
Ideally, the tuning cycle should be long enough to capture the workload characteristics but be as short as possible for better responsiveness.
To balance these two requirements, memory tuning is triggered whenever the accumulated log records exceed the maximum log length.
This allows the memory tuner to capture the workload statistics more accurately by waiting for log-triggered flushes to complete.
For read-heavy workloads, it may take a very long time to produce enough log records.
To address this, the memory tuner also uses a timer-based tuning cycle, e.g., 10 minutes.

\subsection{\revise{Optimality of Memory Tuner}}
\revise{
	\label{sec:tuner-optimality}
The optimality of the memory tuner depends on the shape of $cost(x)$.
To ensure that the memory tuner always finds the global minimum, $cost'(x)$ should have at most one root.
However, after analyzing $cost''(x)$, i.e., the derivative of $cost'(x)$, we have found that this condition may not always hold.
Intuitively, it is easy to see that $write(x)$ is monotonically decreasing and $read_q(x)$ is monotonically increasing.
However, $read_m(x)$ is not monotonic because a larger $x$ may both reduce the number of merge reads and increase the cache miss rate.
Even though the memory tuner may not be able to always find an optimal memory allocation, this does not limit its applicability in practice.
First, it can still find a better memory allocation to reduce the overall I/O cost.
Moreover, we have found that $cost(x)$ for many practical workloads often allows the memory tuner to find the global minimum.
For example, \reffigure{fig:expr-plot-cost} plots the I/O costs for the the YCSB~\cite{ycsb2010} write-heavy workload and the TPC-C~\cite{tpcc} workload.
We used the YCSB write-heavy workload with 50\% writes and 50\% reads.
The operations were distributed among 10 LSM-trees, each of which had 10 million records, following an 80-20 hotspot distribution.
For TPC-C, the scale factor was set at 2000.
(The detailed experimental setup is further described in \refsection{sec:expr-setup}.)
For both workloads, the total I/O cost has one global minimum.
The reason is that the merge read cost $read_m(x)$ is relatively small, even under the YCSB write-heavy workload, because many SSTables at small levels are often cached due to frequent merges and accesses.
Thus, non-monotonicity of $read_m(x)$ does not affect the total I/O cost too much.
We also evaluated other configurations by changing the total memory size and the read-write ratio and found that the total I/O cost always has the same general shape.

\begin{figure}
	\centering
	\includegraphics[width=\linewidth]{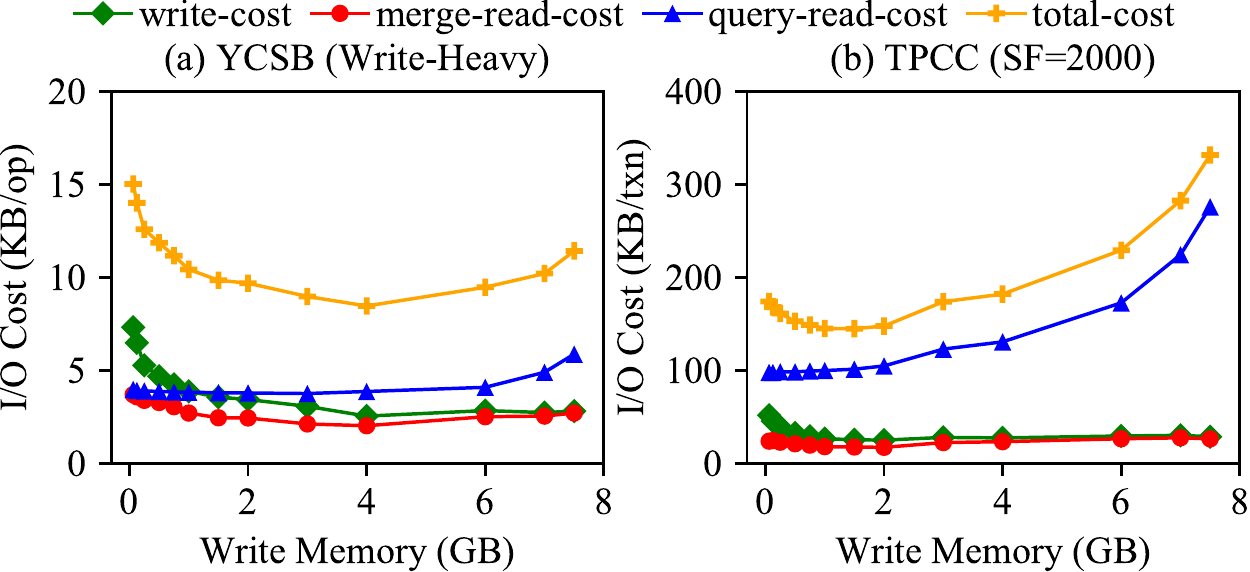}
	\vspace{-0.25in}
	\caption{\revise{I/O Costs under Different Write Memory Sizes}}
	\label{fig:expr-plot-cost}
\end{figure}
}

\section{Experimental Evaluation}
\label{sec:experiment}
In this section, we experimentally evaluate the proposed techniques in the context of Apache AsterixDB~\cite{asterixdb-web}.
Throughout the evaluation, we focus on the following two questions.
First, what are the benefits of the partitioned memory component compared to alternative approaches?
Second, what is the effectiveness of the memory tuner in terms of its accuracy and responsiveness?
In the remainder of this section, we first describe the general experimental setup followed by the detailed evaluation results.

\subsection{Experimental Setup}
\label{sec:expr-setup}
\textbf{Hardware.}
All experiments were performed on a single node m5d.2xlarge on AWS.
The node has an 8-core 2.50GHZ  vCPUs, 32GB of memory, a 300GB NVMe SSD, and a 500GB elastic block store (EBS).
We use the native NVMe for LSM storage and EBS for transactional logging.
The NVMe SSD provides a write throughput of 250MB/s and a read throughput of 500MB/s.
\revise{The EBS also provides a write throughput of 250MB/s.
	Asynchronous log flushing and group commit were further used to ensure that logging on the EBS is not the bottleneck.}
We allocated 26GB of memory for the AsterixDB instance.
Unless otherwise noted, the total storage memory budget, including the buffer cache and the write memory, was set at 20GB.
Both the disk page size and memory page size were set at 16KB.
The maximum transaction log length was set at 10GB.
Finally, we used 8 worker threads to execute workload operations.

\textbf{LSM-tree Setup.}
All LSM-trees used a partitioned leveling merge policy with a size ratio of 10, which is a common setting in existing systems. 
Unless otherwise noted, the number of disk levels was dynamically determined based on the current write memory size.
For the partitioned memory component, its active SSTable size was set at 32MB and the size ratio of the memory merge policy was also set at 10.
We used 2 threads to execute flushes, 2 threads to execute memory merges, and 4 threads to execute disk merges.
In each set of experiments, we first loaded the LSM storage based on the given workload.
Each experiment always started with a fresh copy of the loaded LSM storage.
For both memory and disk levels, we built a Bloom filter for each SSTable with a false positive rate of 1\% to accelerate point lookups.
Finally, both the memory flush threshold and the log truncation threshold were set at 95\%.

\begin{figure*}
	\centering
	\includegraphics[width=1\linewidth]{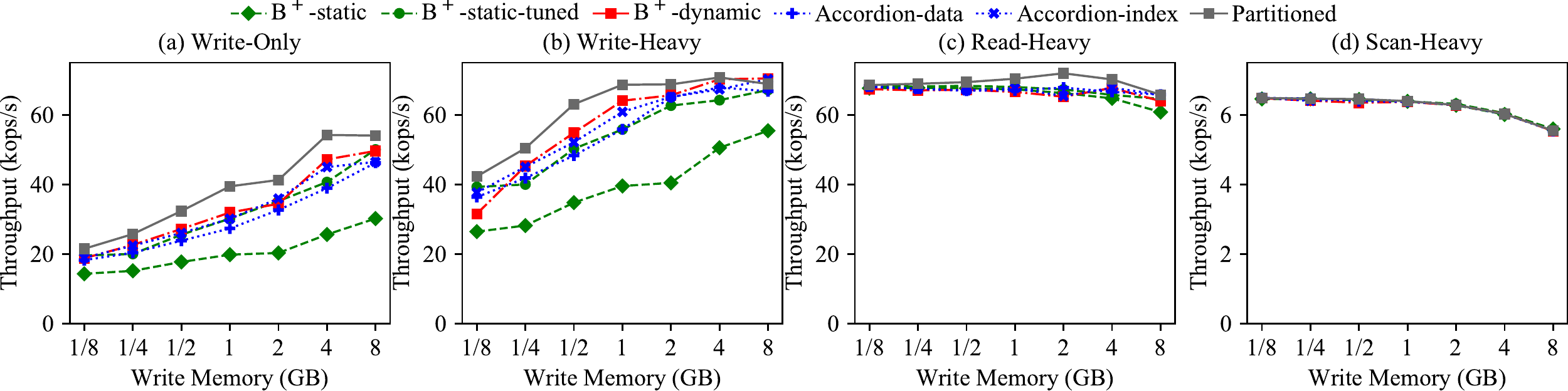}
	\vspace{-0.25in}
	\caption{Experimental Results for a Single LSM-tree}
	\label{fig:expr-singe-lsm}
\end{figure*}

\textbf{Workloads.}
We used two popular benchmarks YCSB~\cite{ycsb2010} and TPC-C~\cite{tpcc}.
YCSB is a popular and extensible benchmark for evaluating key-value stores.
Due to its simplicity, we used YCSB to understand the basic performance of various techniques.
In all experiments, we used the default YCSB record size, where each record has 10 fields with 1KB size in total, and the default Zipfian distribution.
Since YCSB only supports a single LSM-tree, we further extended it to support multiple primary and secondary LSM-trees, which is described in \refsection{sec:expr-write-memory}.
TPC-C is an industrial standard benchmark used to evaluate transaction processing systems.
We chose TPC-C because it represents a more realistic workload with multiple datasets\footnote{\revise{A \emph{dataset} in AsterixDB is equivalent to a \emph{table} in the TPC-C benchmark.}} and secondary indexes.
It should be noted that AsterixDB only supports a basic record-level transaction model without full ACID transactions.
Thus, all transactions in our evaluation were effectively running under the read-uncommitted isolation level from the TPC-C perspective.
Because of this, we disabled the client-triggered aborts (1\%) of the NewOrder transaction.

\subsection{Evaluating Write Memory Management}
\label{sec:expr-write-memory}
We first evaluated the proposed techniques for managing the write memory with the following experiments.
The first set of experiments uses a single LSM-tree to evaluate the basic performance of various memory component structures.
The second set of experiments uses multiple datasets, each of which just has a primary LSM-tree.
The third set of experiments focuses on LSM-based secondary indexes, all belonging to the same dataset.
The last set of experiments uses a more realistic workload that contains multiple primary and secondary indexes.
For the first three sets of experiments, we used the YCSB benchmark~\cite{ycsb2010} due to its simplicity and customizability.
For the last set of experiments, we used the TPC-C benchmark~\cite{tpcc} since it represents a more realistic workload.
\shortonly{Due to space limitations, we leave the third set of experiments to the extended version of this
paper~\cite{lsm-memory-extended}.
In general, we found that the performance trends in the secondary LSM-tree case
were consistent with the results of the multiple primary LSM-tree case.
\revise{~\cite{lsm-memory-extended} also contains further the evaluation of certain design choices, e.g., the choice of flushes and the grouped $L_0$ structure.}}

\hyphenation{data-sets}
\hyphenation{Approach-es}

\textbf{Evaluated Write Memory Management Schemes.}
First, we evaluated two variations of AsterixDB's static memory allocation scheme.
The first variation, called \static, uses AsterixDB's default number of active datasets, which is 8.
The second variation, called \statictuned, configures the number of active datasets parameter setting based on each experiment.
We further evaluated an optimized version of the write memory management scheme (called \dynamic) used in existing systems, e.g., RocksDB and HBase.
\revise{This scheme uses a \btree to manage the memory component of each LSM-tree without any static size limit.}
We also evaluated two variations of Accordion~\cite{accordion2018}.
Accordion separates keys from values by storing keys into an index structure while putting values into a log.
The first variation, called \accordionindex, only merges the indexes without rewriting the logs.
The second variation, called \accordiondata, merges both the indexes and logs.
Finally, we evaluated the partitioned memory component structure, called \partition.
\revise{For both \dynamic and \partition, we evaluated three variations based on the three flush policies described in \refsection{sec:multiple-lsm-tree}, namely max-memory (called \emph{MEM}), min-LSN (called \emph{LSN}), and optimal (called \emph{OPT}).
It should be noted that \dynamic as implemented in existing systems always uses the max-memory policy to flush the LSM-tree with the largest memory component.}

\subsubsection{Single LSM-tree}
In this experiment, the LSM-tree had 100 million records with a 110GB storage size.
We evaluated four types of workloads, namely write-only (100\% writes), write-heavy (50\% writes and 50\% lookups),
read-heavy (5\% writes and 95\% lookups), and scan-heavy (5\% writes and 95\% scans).
A write operation updates an existing key and each scan query accesses a range of 100 records.
Each experiment ran for 30 minutes and the first 10-minute period was excluded when computing the throughput.
It should be noted that in this experiment all flush policies have the identical behavior because there was only one LSM-tree.

\textbf{Basic Performance.}
\reffigure{fig:expr-singe-lsm} shows the throughput of each memory component scheme
under different workloads and write memory sizes.
In general, the write memory mainly impacts write-dominated workloads,
such as write-only and write-heavy, and larger write memory improves the overall throughput by reducing the write cost.
Among these structures, \static always performs the worst since any one LSM-tree is only allocated 1/8 of the write memory.
\dynamic performs slightly better than \statictuned because the former does not leave memory idle by preallocating
two memory components for double buffering.
\partition has the highest throughput under write-dominated workloads since it better utilizes the write memory.
It also improves the overall throughput slightly under the read-heavy workload by reducing write amplification.
For both \dynamic and \partition, the throughput stops increasing after the write memory exceeds 4GB
because flushes are then dominated by log-truncation.
Finally, Accordion does not provide any improvement compared to \dynamic.
\accordiondata actually reduces the overall throughput because a large memory merge will temporarily double the memory usage,
forcing memory components to be flushed.
Moreover, Accordion was designed for reducing GC overhead since HBase~\cite{hbase} uses Java objects to manage memory components.
Although AsterixDB is written in Java, it uses off-heap structures for memory management~\cite{bloat2013,asterixdb-memory}.
In all experiments, its measured GC time was always less than 1\% of the total run time.
Based on these results, and because Accordion is mainly designed for a single LSM-tree,
we excluded Accordion for further evaluation with multiple LSM-trees.

As suggested by \cite{lsm-stability2019}, we further carried out an experiment to evaluate the 99th percentile write latencies of each scheme using a constant data arrival process, whose arrival rate was set at a high utilization level (95\% of the measured maximum write throughput).
We found out that the resulting 99th percentile latencies of all schemes were less than 1s, which suggests that all structures can provide a stable write throughput with a relatively small variance, even under a very high utilization level.

\revise{
\textbf{CPU Overhead of Memory Merges.}
We have seen that \partition outperforms other memory component structures by better utilizing the write memory.
However, it may incur additional CPU overhead due to memory merges.
To evaluate this overhead, we carried out an experiment focusing exclusively on the memory component performance.
We used a smaller YCSB dataset with only 10 million records.
We set the maximum write memory to be 20GB and disabled transaction logging so that the dataset always fits in the memory component.
All operations were executed using a single thread and memory merges were always executed synchronously.
}

\revise{
\reffigure{fig:expr-singe-lsm-mem} shows the resulting throughput under different workloads.
To store the same experiment dataset, \partition only used 12GB of write memory while \dynamic used 15.5GB.
In general, \partition reduces the in-memory throughput by 20\%-40\% as compared to \dynamic due to memory merges,
where the write amplification was about 11.36.
However, it should be noted that in-memory workloads are not the focus of this work.
The partitioned memory component structure thus trades some CPU cycles to reduce the overall disk write amplification.
}

\shortonly{
	\begin{figure}[b]
		\begin{minipage}[t]{0.235\textwidth}
			\centering
			\includegraphics[width=\linewidth]{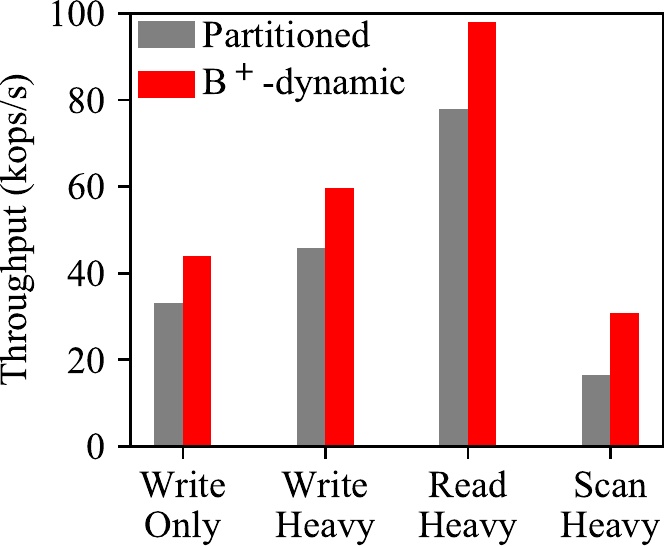}
			\vspace{-0.2in}
			\caption{\revise{Evaluation of Memory Merge Overhead}}
			\label{fig:expr-singe-lsm-mem}
		\end{minipage}
		\hfil
		\begin{minipage}[t]{0.235\textwidth}
			\centering
			\includegraphics[width=\linewidth]{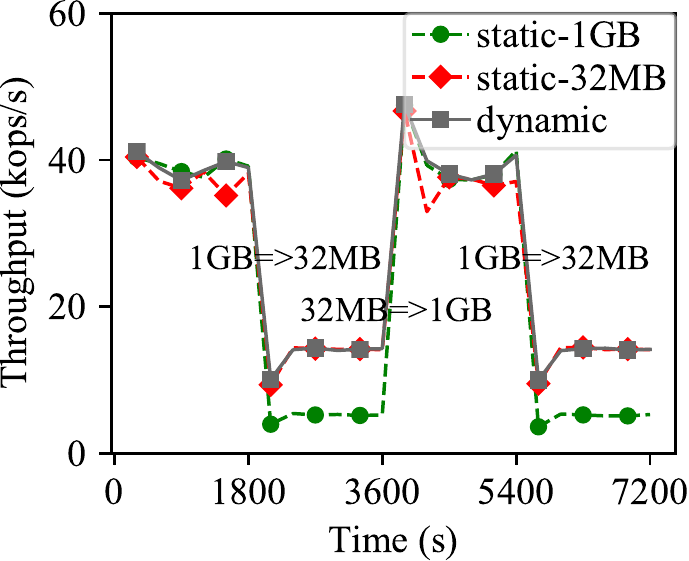}
			\vspace{-0.2in}
			\caption{Write Throughput with Varying Write Memory}
			\label{fig:expr-singe-lsm-dynamic-mem}
		\end{minipage}
	\end{figure}
}

\longonly{
	\begin{figure*}
		\begin{minipage}[t]{0.24\textwidth}
			\centering
			\includegraphics[width=\linewidth]{expr-memory-1}
			\vspace{-0.2in}
			\caption{\revise{Evaluation of Memory Merge Overhead}}
			\label{fig:expr-singe-lsm-mem}
		\end{minipage}
		\hfil
		\begin{minipage}[t]{0.24\textwidth}
			\centering
			\includegraphics[width=\linewidth]{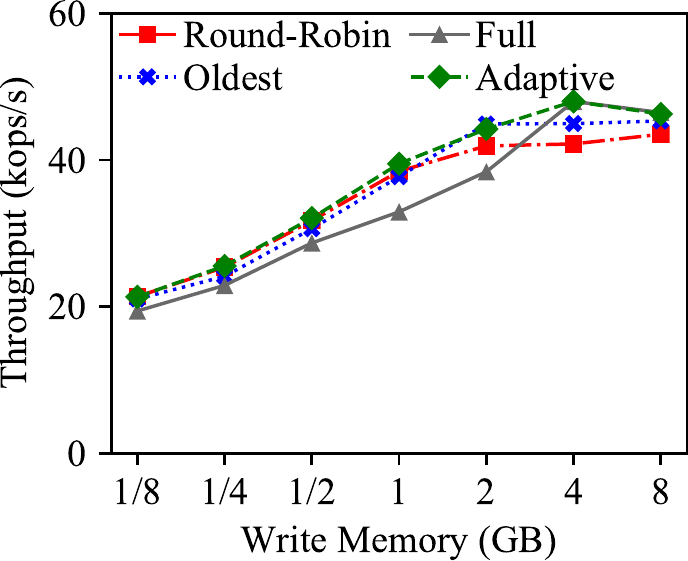}
			\vspace{-0.2in}
			\caption{\revise{Evaluation of Flush Heuristics}}
			\label{fig:expr-single-flush}
		\end{minipage}
		\hfil
		\begin{minipage}[t]{0.24\textwidth}
			\centering
			\includegraphics[width=\linewidth]{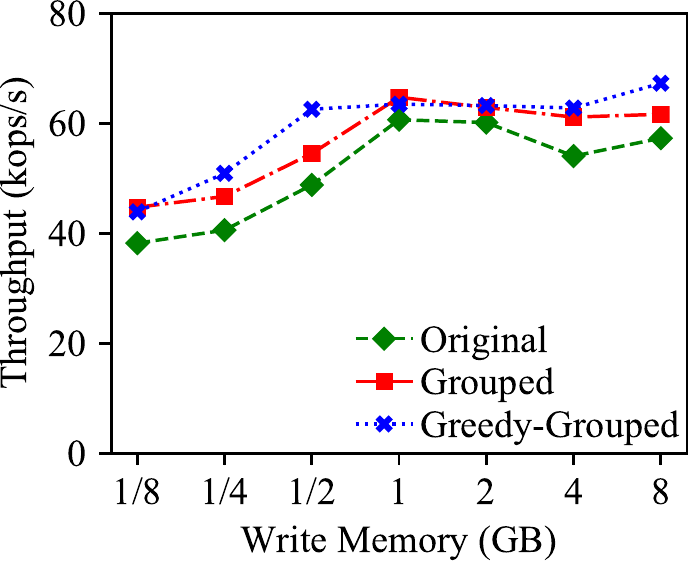}
			\vspace{-0.2in}
			\caption{\revise{Evaluation of L0 Structure}}
			\label{fig:expr-single-l0}
		\end{minipage}
		\hfil
		\begin{minipage}[t]{0.24\textwidth}
			\centering
			\includegraphics[width=\linewidth]{expr-single-dynamic}
			\vspace{-0.2in}
			\caption{Write Throughput with Varying Write Memory}
			\label{fig:expr-singe-lsm-dynamic-mem}
		\end{minipage}
	\end{figure*}
}

\longonly{
\revise{
\textbf{Evaluation of Flush Heuristics.}
In \refsection{sec:partitioned-memory-component}, we described a number of heuristics to adaptive select SSTables to flush by distinguishing
memory-triggered flushes, log-triggered flushes, and full flushes.
To evaluate the benefit of the proposed heuristics, we performed an experiment with the following flush strategies.
The first strategy, called \emph{Round-Robin}, always flushes SSTables at the last level in a round-robin way.
The second strategy, called \emph{Oldest}, always flushes the oldest SSTable along with all overlapping SSTables at higher levels.
The third strategy, called \emph{Full}, always performs full flushes by flushing all SSTables together.
Finally, the proposed strategy, called \emph{Adaptive}, combines different flush strategies as described in \refsection{sec:partitioned-memory-component}.
We used the write-only workload since different flush strategies mainly impact the write performance.

\reffigure{fig:expr-single-flush} shows the resulting write throughput under different write memory sizes.
As one can see, there is no clear winner among the first three strategies.
Under small write memory (smaller than 1GB) with many memory-triggered flushes, \emph{Round-Robin} achieves the highest throughput.
\emph{Oldest} performs better when the write memory size becomes larger with many log-triggered flushes,
as flushing the oldest SSTable better facilitates log truncation.
When the write memory size becomes even larger (larger than 4GB), log-triggered flushes become dominating and thus \emph{Full} achieves the best throughput.
This experiment also shows that by combining these three flush strategies together, the proposed flush strategy always achieves
the best possible write throughput.
}
}

\longonly{
\revise{
\textbf{Evaluation of Grouped $L_0$ Structure.}
To evaluate the effectiveness of the grouped $L_0$ structure that organizes non-overlapping SSTables into groups,
we performed an experiment with the following alternative structures.
First, we evaluated the $L_0$ structure in the original LSM-tree design shown in \reffigure{fig:partitioned-lsm}, called \emph{Original}, which simply tolerates $N$ SSTables.
The second structure, called \emph{Grouped}, organizes SSTables into groups
but does not use the proposed heuristics to minimize the write amplification.
Specifically, this structure always picks the leftmost SSTable from the oldest group to merge, along with overlapping SSTables from newer groups.
The last structure, called \emph{Greedy-Grouped}, further uses the proposed heuristics to minimize the write amplification.
Since all these $L_0$ structures have the same worst-case query performance, we used a write-only workload to evaluate their impact on writes.
To show the impact of the alternative $L_0$ structures, we set the maximum number of groups at $L_0$ as 4.

\reffigure{fig:expr-single-l0} shows the resulting write throughput of alternative $L_0$ structures.
The $L_0$ structure used in the original LSM-tree design led to the worst write throughput because it did not exploit the disjoint SSTables to reduce the write cost.
By organizing disjoint SSTables into groups, the grouped $L_0$ structure increased the write throughput since multiple overlapping SSTables can be merged at once.
Finally, the proposed greedy heuristics further increased the write throughput by greedily selecting SSTables to merge to minimize the write amplification.
}
}

\textbf{Benefits of Dynamically Adjusting Disk Levels.}
To evaluate the benefit of dynamically adjusting disk levels as the write memory changes,
we conducted an experiment where the write memory size
alternates between 1GB and 32MB every 30 minutes.
Each experiment ran for two hours in total.
We used the partitioned memory component structure but the disk levels were determined differently.
In addition to the proposed approach that adjusts disk levels dynamically (called \emph{dynamic}), we used two baselines where the number of disk levels is determined statically by assuming that the write memory is always 32MB (called \emph{static-32MB}) or always 1GB (called \emph{static-1GB}).
The resulting write throughput, aggregated over 5-minute windows, is shown in \reffigure{fig:expr-singe-lsm-dynamic-mem}.
The dynamic approach always has the highest throughput, which confirms the utility of adjusting disk levels as the write memory changes.
Moreover, we see that having fewer levels when the write memory is small has a more negative impact than having more levels when the write memory is large
since the write throughput for \emph{static-1GB} is much lower under the small write memory.

\begin{figure}
	\centering
	\includegraphics[width=\linewidth]{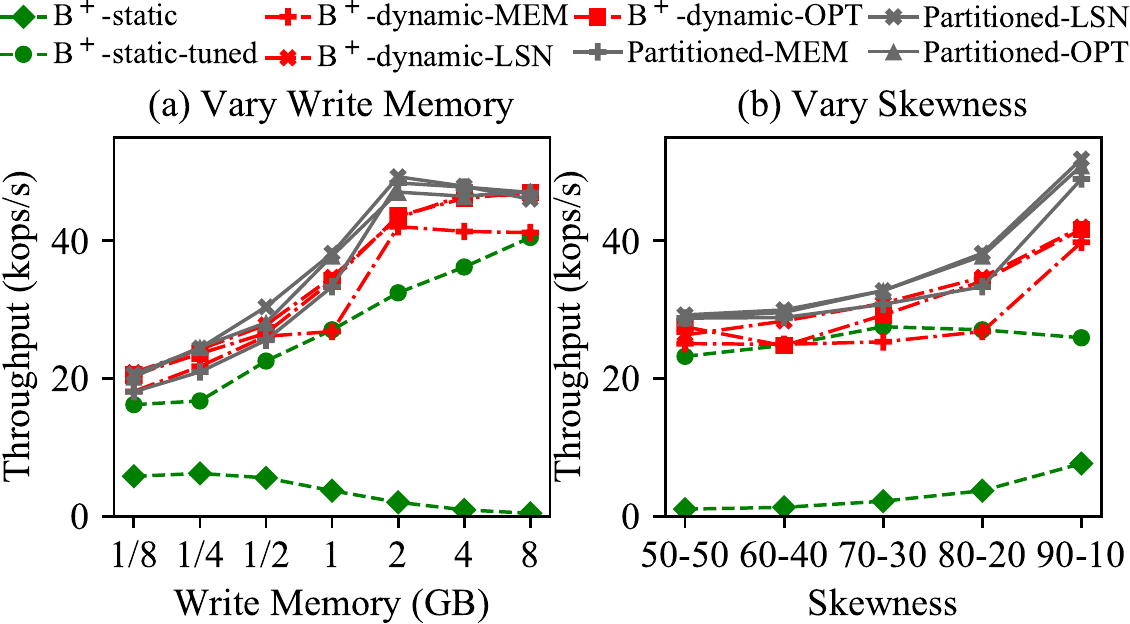}
	\vspace{-0.2in}
	\caption{\revise{Evaluation of Multiple Primary LSM-trees}}
	\label{fig:expr-multi-primary-lsm}
\end{figure}

\longonly{
	\begin{figure*}
	\centering
	\includegraphics[width=0.85\linewidth]{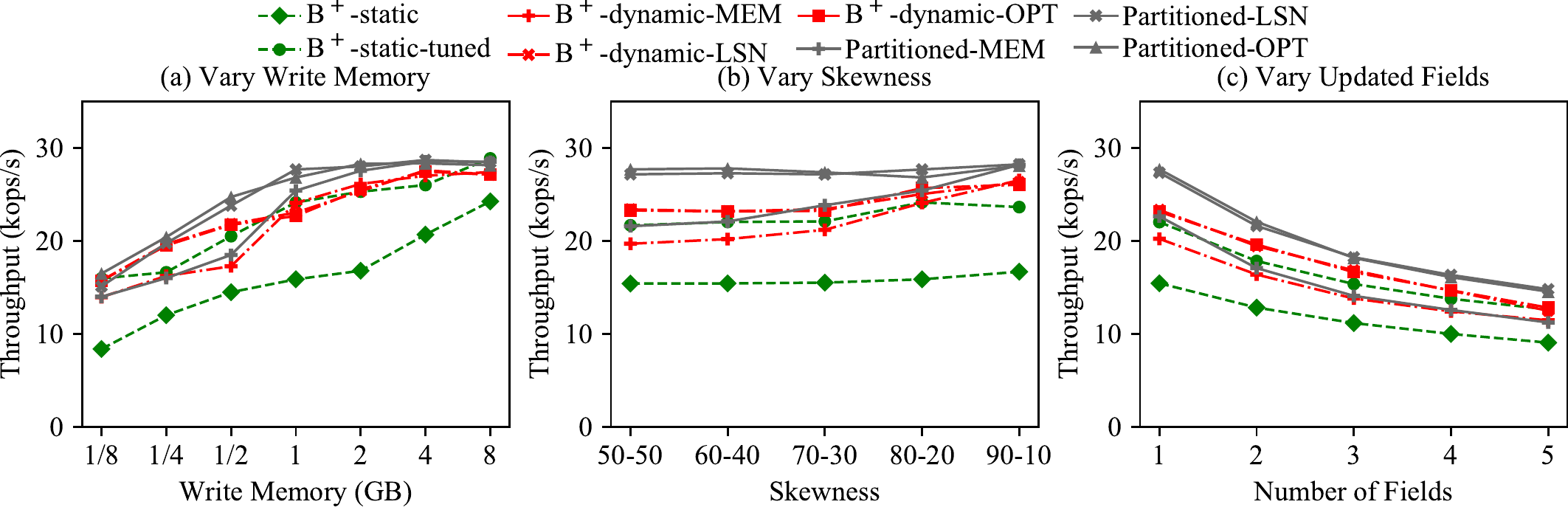}
	\vspace{-0.1in}
	\caption{Evaluation of Multiple Secondary LSM-trees}
	\label{fig:expr-multi-secondary-lsm}
\end{figure*}
}

\begin{figure*}
	\centering
	\includegraphics[width=\linewidth]{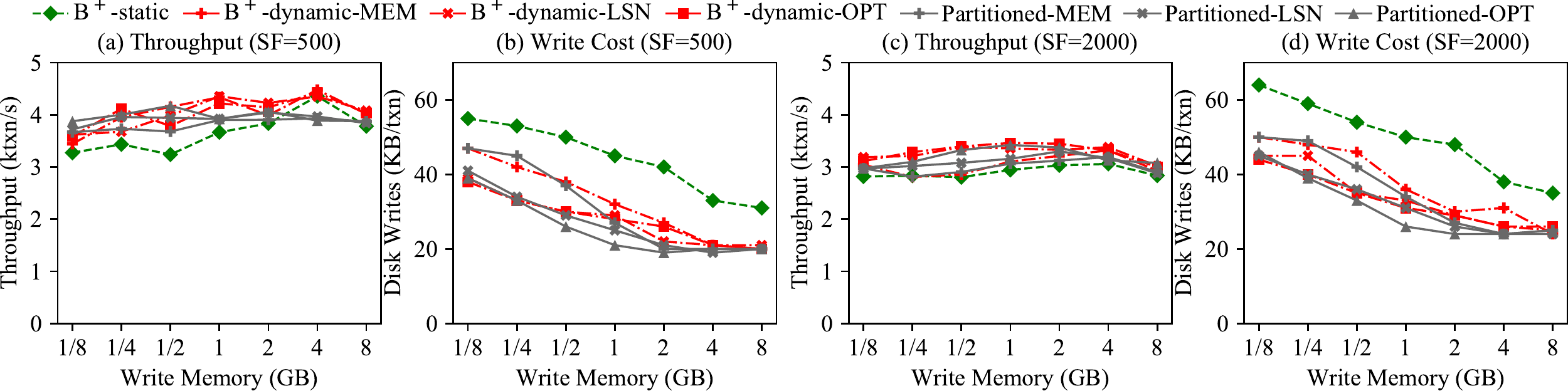}
	\vspace{-0.25in}
	\caption{\revise{Experimental Results on TPC-C}}
	\label{fig:expr-tpcc}
\end{figure*}

\subsubsection{Multiple Primary LSM-trees}
\label{sec:expr-mult-lsm}
In this set of experiments, we used 10 primary LSM-trees, each of which had 10 million records.
Since the write memory mainly impacts write performance, a write-only workload was used in this experiment.
Writes were distributed among the multiple LSM-trees following a hotspot distribution,
where $x\%$ of the writes go to $y\%$ of the LSM-trees.
For example, an 80-20 distribution means that 80\% of the writes go to 20\% of the LSM-trees, i.e., 2 hot LSM-trees,
while the 20\% of the writes go to 80\% of the LSM-trees, yielding 8 cold LSM-trees.
Within each LSM-tree, writes still followed YCSB's default Zipfian distribution.

\textbf{Impact of Write Memory.}
We first evaluated the impact of the write memory size by fixing the skewness to be 80-20.
The resulting write throughput is shown in \reffigure{fig:expr-multi-primary-lsm}a.
Note that \static results in a much lower throughput because of thrashing.
Since the default number of active datasets in AsterixDB is only 8, some LSM-trees have to be constantly activated and deactivated,
resulting in many tiny flushes.
\statictuned avoids the thrashing problem, but it still performs worse than
the other baselines because it does not differentiate hot LSM-trees from cold ones.
\revise{Both \dynamic and \partition allocate the write memory dynamically, improving the write throughput.
Moreover, we see that the min-LSN and optimal flush policies perform better than the max-memory flush policy for both structures.}
Since the max-memory policy always flushes the largest memory component, the memory components of the cold LSM-trees are not flushed until they are large enough or until the transaction log has to be truncated.
The min-LSN policy also has a write throughput comparable to the optimal policy,
which makes it a good approximation but with less implementation complexity.
\revise{Finally, even under the same flush policy, we see that \partition still outperforms \dynamic because it performs memory merges
to further increase memory utilization.
}

\textbf{Impact of Skewness.}
Next, we evaluated the impact of skewness by fixing the write memory to be 1GB.
The resulting write throughput is shown in \reffigure{fig:expr-multi-primary-lsm}b.
All memory component structures except \statictuned benefit from skewed workloads.
The problem of is that \statictuned always allocates the write memory evenly to the active datasets
without differentiating hot LSM-trees from cold ones.
For \static, the thrashing problem is alleviated under skewed workloads since most writes go to a small number of LSM-trees.
\revise{When the workload is more skewed, we see two interesting trends.
First, under the min-LSN and optimal flush policies, the performance difference between \partition and \dynamic becomes larger.
This is because a small number of hot LSM-trees occupy most of the write memory, allowing more memory merges to be performed in these hot LSM-trees
to reduce the write amplification.
Moreover, the performance differences among the three flush policies also become larger
since the min-LSN and optimal policies allocate more write memory to the hot LSM-trees.
}

\longonly{
\subsubsection{Multiple Secondary LSM-trees}
We further evaluated the alternative memory component structures using multiple secondary LSM-trees for one dataset.
The dataset had one primary LSM-tree and 10 secondary LSM-trees, with one secondary LSM-tree per field.
The primary LSM-tree had 50 million records with 55GB storage size, and each secondary LSM-tree was about 5GB.
As before, we used the write-only workload to focus on write performance.
It should be noted that each write must also performs a primary index lookup to cleanup secondary indexes~\cite{lsm-storage2019}.
Unless otherwise noted, each write only updates one secondary field,
but the choice of updated fields followed the same hotspot distribution as in \refsection{sec:expr-mult-lsm}.
For each field, its values followed the default Zipfian distribution used in YCSB.

\textbf{Impact of Write Memory.}
First, we varied the total write memory to evaluate its impact on the different memory component structures with secondary indexes.
The resulting write throughput is shown in \reffigure{fig:expr-multi-secondary-lsm}a.
In general, the results are consistent with the multiple primary LSM-tree case in \reffigure{fig:expr-multi-primary-lsm}a.
Note that \statictuned achieves better performance in this case because \statictuned allocates the write memory at a dataset level.
Thus, the primary LSM-tree and all secondary LSM-trees share the same memory budget, which is similar to other schemes.

\textbf{Impact of Updated Field Skewness.}
We further varied the skewness of updated fields to study its impact on write throughput.
\reffigure{fig:expr-multi-secondary-lsm}b shows the resulting write throughput.
As one can see, the skewness of updated fields has a smaller performance impact here compared to 
the multiple primary LSM-tree case in \reffigure{fig:expr-multi-primary-lsm}b
because the size of a secondary LSM-tree is much smaller than the primary one.
Moreover, \dynamicmem and \partitionmem, both of which used the max-memory flush policy, still benefit from a more skewed workload.
The reason is that when most writes access a small number of hot secondary indexes, 
the size of their memory components will grow faster and they will be selected by the max-memory policy to flush.

\textbf{Impact of Number of Updated Fields.}
Finally, we studied the performance impact of the number of updated fields per write, ranging from 1 to 5.
The resulting throughput is shown in \reffigure{fig:expr-multi-secondary-lsm}c.
Increasing the number of updated fields per write negatively impacts the write throughput because each logical write produces more physical writes.
Because of this, the write throughput of all memory component structures decreases in the same way when each write updates more fields.

}
\begin{figure*}
	\centering
	\includegraphics[width=\linewidth]{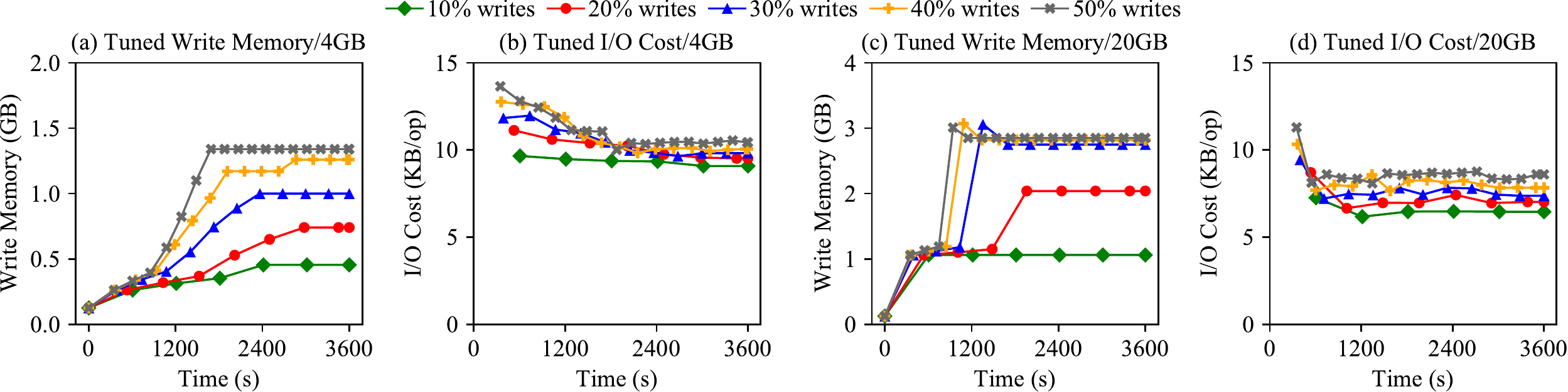}
	\vspace{-0.25in}
	\caption{Evaluation of Memory Tuner on YCSB}
	\label{fig:expr-tuner-ycsb}
\end{figure*}

\subsubsection{TPC-C Results}
Finally, we used the TPC-C benchmark to evaluate the alternative memory management schemes on a more realistic workload.
We used two scale factors (SF) of TPC-C, i.e., 500, which results in a 50GB storage size, and 2000, which results in a 200GB storage size.
Each experiment ran for one hour and the throughput was measured excluding the first 30 minutes.

The resulting throughput and the per-transaction disk writes (KB) under the two scale factors are shown in \reffigure{fig:expr-tpcc}.
Note that \statictuned is omitted here, because the number of active datasets in TPC-C is 8, which is the same as the default value used in AsterixDB.
\static still has the highest I/O cost because it allocates write memory evenly to all datasets.
TPC-C contains some hot datasets, such as order\_line and stock, that receive most of the writes, as well as some cold datasets, such as warehouse and district, that only require a few megabytes of write memory.
\revise{As we have seen in \reffigure{fig:expr-multi-primary-lsm}, the min-LSN and optimal policies have reduced the write cost for both \dynamic and \partition.
\partitionopt also led to the lowest write cost via extra memory merges, improving the system I/O efficiency.
However, since TPC-C is a CPU-heavy workload, doing so may not always improve the overall transaction throughput due to the CPU overhead of memory merges as we have seen before.
When the workload is CPU-bound at scale factor 500, the extra CPU overhead incurred by memory merges actually decreases
the overall throughput as compared to \dynamic.}
Thus, we observe that it is useful to design a memory management scheme to balance the CPU overhead and the I/O cost, which we leave as future work.
Finally, the results also show that increasing the write memory may not always increase the overall transaction throughput.
For example, when the scale factor is 2000, the optimal throughput is reached when the write memory is between 1GB and 2GB.
This confirms the importance of memory tuning, which will be evaluated next.

\subsubsection{Summary}
As all experiments have illustrated, it is important to utilize a large write memory efficiently to reduce the I/O cost.
Although the static memory allocation scheme is relatively simple and robust,
it leads to sub-optimal performance because the write memory is always evenly allocated to active datasets.
The optimized version of the memory management scheme used by existing systems
reduces the I/O cost by dynamically allocating the write memory to active LSM-trees.
This still does not achieve optimal performance, however, because it fails to manage large memory components efficiently
and its choice of flushes does not optimize the overall write cost.
Finally, the proposed partitioned memory component structure and the optimal flush policy minimize the write cost for all workloads.
The use of partitioned memory components manages the large write memory more effectively to reduce the write amplification of a single LSM-tree.
The optimal flush policy allocates the write memory to multiple LSM-trees based on their write rates to minimize the overall write cost.
However, the partitioned memory component structure may incur extra CPU overhead, which makes it less suitable for CPU-heavy workloads.
Finally, we have observed that the min-LSN policy achieves comparable performance to the optimal policy,
which makes it a good approximation but with less implementation complexity.

\subsection{Evaluating the Memory Tuner}
We now proceed to evaluate the memory tuner with the focus on the following questions:
First, what are the basic mechanics of the memory tuner \revise{in terms of how it tunes the memory allocation for different workloads}?
Second, what is the accuracy of the memory tuner as compared to manually tuned memory allocation?
Finally, how responsive is the memory tuner when the workload changes?

In all experiments below, the initial write memory size was set at 64MB and the simulated cache size was set to 128MB.
Unless otherwise noted, other settings of the memory tuner, such as the number of samples for fitting the linear function,
the stopping threshold, and the maximum step size, all used the default values given in \refsection{sec:memory-tuner-finding}.

\subsubsection{Basic Mechanics}
\revise{To understand how the memory tuner performs memory tuning to reduce the I/O cost for different workloads},
we carried out a set of experiments using YCSB~\cite{ycsb2010} with a single LSM-tree.
\revise{We set both weights $\omega$ and $\gamma$ to 1 since we focus on the I/O cost in this experiment.}
The LSM-tree had 100 million records with 110GB in total.
We used a mixed read/write workload where the write ratio varied from 10\% to 50\%.
The total memory budget was set at 4GB or 20GB.
Each experiment ran for 1 hour.

The tuned write memory size and the corresponding I/O costs over time are shown in \reffigure{fig:expr-tuner-ycsb}.
Note that each point denotes one tuning step performed by the memory tuner.
We see that the memory tuner balances the relative gain of allocating more memory to the write memory and the buffer cache
to \revise{reduce} the overall I/O cost.
As shown in Figures \ref{fig:expr-tuner-ycsb}a and \ref{fig:expr-tuner-ycsb}c,
when the overall memory budget is fixed, the memory tuner allocates more write memory when the write ratio is increased
because the benefit of having a large write memory increases.
Moreover, by comparing the allocated write memory sizes in Figures \ref{fig:expr-tuner-ycsb}a and \ref{fig:expr-tuner-ycsb}c,
we can see that when the write ratio is fixed, the memory tuner also allocates more write memory when the total memory becomes larger.
This is because the benefit of having more buffer cache memory plateaus.
Finally, as shown in Figures \ref{fig:expr-tuner-ycsb}b and \ref{fig:expr-tuner-ycsb}d,
the overall I/O cost also decreases after the memory allocation is tuned over time.

\begin{figure}
	\centering
	\includegraphics[width=\linewidth]{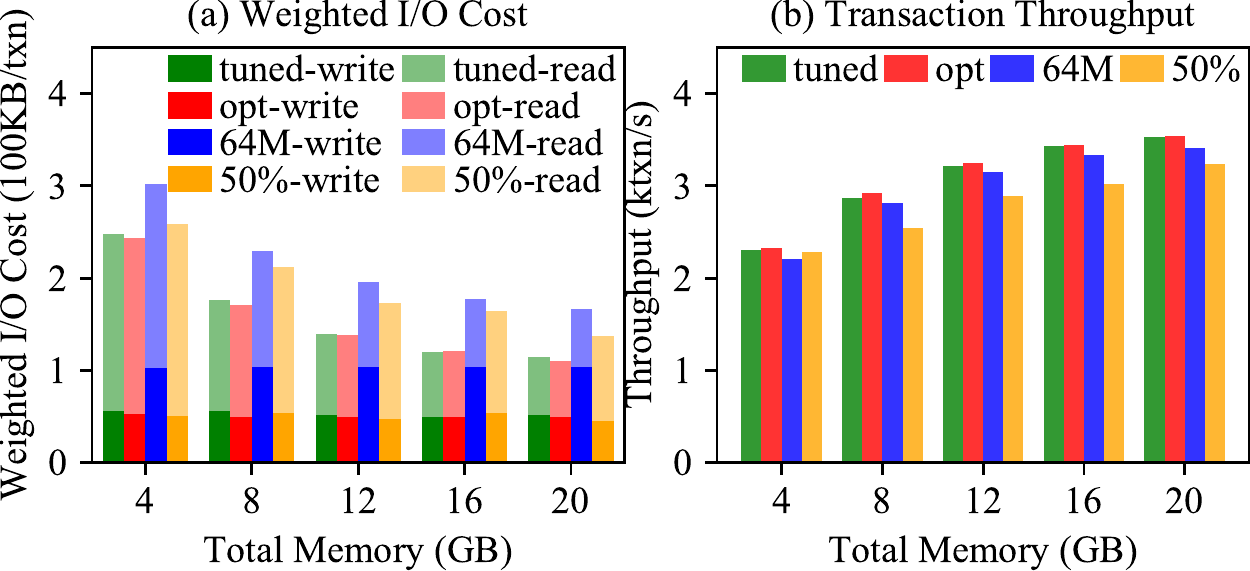}
	\vspace{-0.25in}
	\caption{\revise{Memory Tuner's Accuracy on TPC-C}}
	\label{fig:expr-tuner-tpcc-accuracy}
\end{figure}

\begin{figure}[b]
	\centering
	\includegraphics[width=\linewidth]{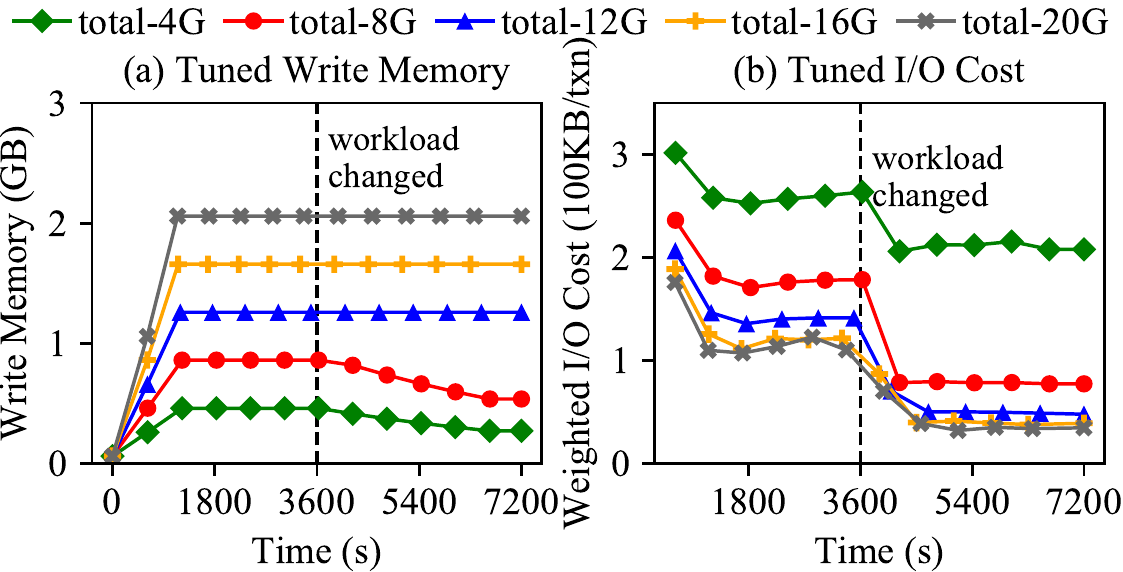}
	\vspace{-0.25in}
	\caption{\revise{Memory Tuner's Responsiveness on TPC-C}}
	\label{fig:expr-tuner-tpcc-response}
\end{figure}

\subsubsection{Accuracy}
To evaluate the accuracy of the memory tuner, we carried out a set of experiments on TPC-C
to compare the tuned performance versus the optimal performance.
Here we used TPC-C because it represents a more complex and more realistic workload than YCSB.
The scale factor was set at 2000.
\revise{Since for our SSD writes are twice as expensive as reads, we set the write weight $\omega$ to be $2$ and the read weight $\gamma$
	to be $1$ in the remaining experiments to balance these two costs.}
\revise{To find the optimal memory allocation (called \emph{opt}), we used an exhaustive search to evaluate different memory allocations with an increment of 128MB.}
To show the effectiveness of the memory tuner, we included two additional baselines.
The first baseline (called \emph{64M}) always set the write memory at 64MB, which was the starting point of the memory tuner.
The second baseline (called \emph{50\%}) divided the total memory budget evenly between the buffer cache and the write memory.
We further varied the total memory budget from 4GB to 20GB.
Each experiment ran for 1 hour and the initial 30 minutes were excluded from the measurement.

\revise{\reffigure{fig:expr-tuner-tpcc-accuracy} shows the weighted I/O cost per transaction
and the transaction throughput for the different memory allocations.
Using exhaustive search, we found that minimizing the weighted I/O cost also maximized the transaction throughput.
The auto-tuned I/O cost and throughput (called \emph{tuned}) are very close to the optimal ones found via exhaustive search,
which shows the effectiveness of our memory tuner.}
Moreover, the memory tuner performs notably better than the two heuristic-based baselines.
Allocating a small write memory minimizes the read cost but leads to a higher write cost.
In contrast, allocating a large write memory minimizes the write cost but the read cost becomes much higher.
\revise{As a result, both allocations also fail to maximize the overall transaction throughput.}

\subsubsection{Responsiveness}
Finally, we used a variation of TPC-C to evaluate the responsiveness of the memory tuner.
This experiment started with the default TPC-C transaction mix and the workload changed into a read-mostly variation,
one which contains 5\% write transactions, i.e., new\_order, payment, and delivery, and 95\% read transactions, i.e., order\_status and stock\_level.
Each experiment ran for two hours and the workload was changed after the first hour.
The resulting allocated write memory and \revise{weighted I/O cost} over time are shown in \reffigure{fig:expr-tuner-tpcc-response}.
After the workload changes, the memory tuner immediately starts to allocate more memory to the buffer cache.
Note that the write memory decreases relatively slowly because the memory tuner limits its step size to 10\% of the current write memory size
to ensure stability.
However, this does not impact the overall I/O cost too much because the buffer cache already occupies most of the memory.
Also note that the write memory size does not change in response to the workload shift when the total memory is larger than 8GB.
This is because the buffer cache already occupies most of the memory and allocating more memory would not change the I/O cost too much.

\shortonly{
	In~\cite{lsm-memory-extended}, we further evaluated the impact of the maximum step size on the responsiveness and stability of the memory tuner.
	Even though a large step size allows the memory tuner to change the memory allocation more rapidly, it also leads to instability and oscillation.
	Thus, the memory tuner's default maximum step size is set at 10\% to ensure stability while providing reasonable responsiveness.
}

\longonly{
To study the impact of the maximum step size on the responsiveness and stability of the memory tuner,
we further carried out an experiment that varies the maximum step size from 10\% to 100\%.
The total memory was set at 12GB.
Each experiment ran for four hours and the workload changed from the the default TPC-C mix into the read-heavy mix after the first hour.
The tuned write memory and \revise{weighted I/O cost} over time are shown in Figures \ref{fig:expr-tuner-tpcc-response-step}a and \ref{fig:expr-tuner-tpcc-response-step}b respectively.
As the results show, increasing the maximum step size improves responsiveness by allowing the memory tuner to change the memory allocation more quickly.
However, this also negatively impacts the memory tuner's stability and leads to some oscillation.
Also note that decreasing the write memory more rapidly has a very small impact on the I/O cost since the buffer cache already occupies most of the memory.
Thus, the memory tuner's default maximum step size is set at 10\% to ensure stability while providing reasonable responsiveness.

\begin{figure}
	\centering
	\includegraphics[width=\linewidth]{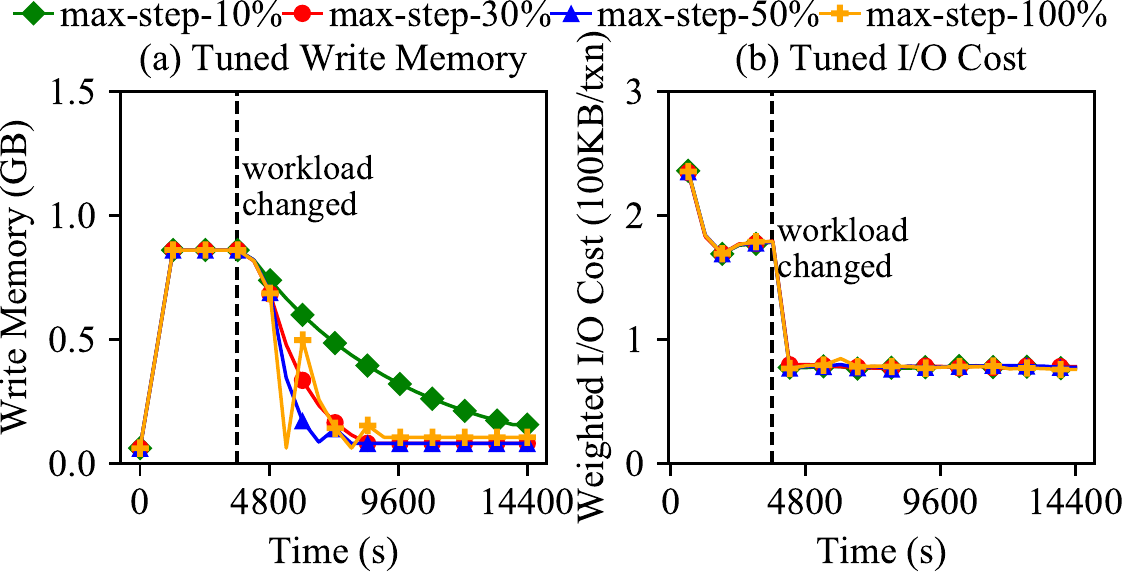}
	\vspace{-0.2in}
	\caption{Impact of Maximum Step Size on Memory Tuner's Responsiveness}
	\label{fig:expr-tuner-tpcc-response-step}
\end{figure}
}

\subsubsection{Summary}
We have evaluated the memory tuner in terms of its mechanics, accuracy, and responsiveness.
Our tuner uses a white-box to minimize the overall I/O cost
based on the relative gains of allocating more memory to the buffer cache or to the write memory.
The experimental results show that this white-box approach enables the memory tuner to achieve both high accuracy with reasonable responsiveness,
making it suitable for online tuning.

\section{Conclusion}
\label{sec:conclusion}
In this paper, we have described and evaluated a number of techniques to break down the memory walls in LSM-based storage systems.
We first presented an LSM memory management architecture that facilitates adaptive memory management.
We further proposed a partitioned memory component structure with new flush polices to better utilize the write memory
in order to minimize the overall write cost.
To break down the memory wall between the write memory and the buffer cache,
we further introduced a memory tuner that uses a white-box approach to continuously tune the memory allocation.
We have empirically demonstrated that
these techniques together enable adaptive memory management to minimize the I/O cost for LSM-based storage systems.
\revise{In the future, we plan to extend the memory tuner to incorporate other memory regions, such as query operator memory, to perform global memory tuning.}

\begin{acks}
	We thank Mark Callaghan and anonymous reviewers for their helpful comments. This work has been supported by NSF awards CNS-1305430, IIS-1447720, IIS-1838248, and CNS-1925610 along with industrial support from Amazon, Google, and Microsoft and support from the Donald Bren Foundation (via a Bren Chair).
\end{acks}

\shortonly{
\newpage
}
\balance

\bibliographystyle{ACM-Reference-Format}
\bibliography{lsm}


\begin{thebibliography}{62}


\ifx \showCODEN    \undefined \def \showCODEN     #1{\unskip}     \fi
\ifx \showDOI      \undefined \def \showDOI       #1{#1}\fi
\ifx \showISBNx    \undefined \def \showISBNx     #1{\unskip}     \fi
\ifx \showISBNxiii \undefined \def \showISBNxiii  #1{\unskip}     \fi
\ifx \showISSN     \undefined \def \showISSN      #1{\unskip}     \fi
\ifx \showLCCN     \undefined \def \showLCCN      #1{\unskip}     \fi
\ifx \shownote     \undefined \def \shownote      #1{#1}          \fi
\ifx \showarticletitle \undefined \def \showarticletitle #1{#1}   \fi
\ifx \showURL      \undefined \def \showURL       {\relax}        \fi
\providecommand\bibfield[2]{#2}
\providecommand\bibinfo[2]{#2}
\providecommand\natexlab[1]{#1}
\providecommand\showeprint[2][]{arXiv:#2}

\bibitem[\protect\citeauthoryear{??}{ast}{[n.d.]}]%
        {asterixdb-web}
 \bibinfo{year}{[n.d.]}\natexlab{}.
\newblock \bibinfo{title}{{AsterixDB}}.
\newblock \bibinfo{howpublished}{\url{https://asterixdb.apache.org/}}.
\newblock


\bibitem[\protect\citeauthoryear{??}{cas}{[n.d.]}]%
        {cassandra}
 \bibinfo{year}{[n.d.]}\natexlab{}.
\newblock \bibinfo{title}{Cassandra}.
\newblock \bibinfo{howpublished}{\url{http://cassandra.apache.org/}}.
\newblock


\bibitem[\protect\citeauthoryear{??}{hba}{[n.d.]}]%
        {hbase}
 \bibinfo{year}{[n.d.]}\natexlab{}.
\newblock \bibinfo{title}{{HBase}}.
\newblock \bibinfo{howpublished}{\url{https://hbase.apache.org/}}.
\newblock


\bibitem[\protect\citeauthoryear{??}{lev}{[n.d.]}]%
        {leveldb}
 \bibinfo{year}{[n.d.]}\natexlab{}.
\newblock \bibinfo{title}{{LevelDB}}.
\newblock \bibinfo{howpublished}{\url{http://leveldb.org/}}.
\newblock


\bibitem[\protect\citeauthoryear{??}{myr}{[n.d.]}]%
        {myrocks}
 \bibinfo{year}{[n.d.]}\natexlab{}.
\newblock \bibinfo{title}{{MyRocks}}.
\newblock \bibinfo{howpublished}{\url{https://http://myrocks.io/}}.
\newblock


\bibitem[\protect\citeauthoryear{??}{roc}{[n.d.]}]%
        {rocksdb}
 \bibinfo{year}{[n.d.]}\natexlab{}.
\newblock \bibinfo{title}{{RocksDB}}.
\newblock \bibinfo{howpublished}{\url{http://rocksdb.org/}}.
\newblock


\bibitem[\protect\citeauthoryear{??}{tpc}{[n.d.]}]%
        {tpcc}
 \bibinfo{year}{[n.d.]}\natexlab{}.
\newblock \bibinfo{title}{{TPC-C}}.
\newblock \bibinfo{howpublished}{\url{http://www.tpc.org/tpcc/}}.
\newblock


\bibitem[\protect\citeauthoryear{Agrawal, Chaudhuri, Kollar, Marathe,
  Narasayya, and Syamala}{Agrawal et~al\mbox{.}}{2005}]%
        {sqlserver2005}
\bibfield{author}{\bibinfo{person}{Sanjay Agrawal}, \bibinfo{person}{Surajit
  Chaudhuri}, \bibinfo{person}{Lubor Kollar}, \bibinfo{person}{Arun Marathe},
  \bibinfo{person}{Vivek Narasayya}, {and} \bibinfo{person}{Manoj Syamala}.}
  \bibinfo{year}{2005}\natexlab{}.
\newblock \showarticletitle{Database tuning advisor for {Microsoft SQL Server}
  2005}. In \bibinfo{booktitle}{\emph{ACM International Conference on
  Management of Data (SIGMOD)}}. ACM, \bibinfo{pages}{930--932}.
\newblock


\bibitem[\protect\citeauthoryear{Alsubaiee, Altowim, Altwaijry, Behm, Borkar,
  Bu, Carey, Cetindil, Cheelangi, Faraaz, Gabrielova, Grover, Heilbron, Kim,
  Li, Li, Ok, Onose, Pirzadeh, Tsotras, Vernica, Wen, and Westmann}{Alsubaiee
  et~al\mbox{.}}{2014a}]%
        {asterixdb2014}
\bibfield{author}{\bibinfo{person}{Sattam Alsubaiee}, \bibinfo{person}{Yasser
  Altowim}, \bibinfo{person}{Hotham Altwaijry}, \bibinfo{person}{Alexander
  Behm}, \bibinfo{person}{Vinayak Borkar}, \bibinfo{person}{Yingyi Bu},
  \bibinfo{person}{Michael Carey}, \bibinfo{person}{Inci Cetindil},
  \bibinfo{person}{Madhusudan Cheelangi}, \bibinfo{person}{Khurram Faraaz},
  \bibinfo{person}{Eugenia Gabrielova}, \bibinfo{person}{Raman Grover},
  \bibinfo{person}{Zachary Heilbron}, \bibinfo{person}{Young-Seok Kim},
  \bibinfo{person}{Chen Li}, \bibinfo{person}{Guangqiang Li},
  \bibinfo{person}{Ji~Mahn Ok}, \bibinfo{person}{Nicola Onose},
  \bibinfo{person}{Pouria Pirzadeh}, \bibinfo{person}{Vassilis Tsotras},
  \bibinfo{person}{Rares Vernica}, \bibinfo{person}{Jian Wen}, {and}
  \bibinfo{person}{Till Westmann}.} \bibinfo{year}{2014}\natexlab{a}.
\newblock \showarticletitle{{AsterixDB}: A Scalable, Open Source {BDMS}}.
\newblock \bibinfo{journal}{\emph{Proceedings of the VLDB Endowment (PVLDB)}}
  \bibinfo{volume}{7}, \bibinfo{number}{14} (\bibinfo{year}{2014}),
  \bibinfo{pages}{1905--1916}.
\newblock


\bibitem[\protect\citeauthoryear{Alsubaiee, Behm, Borkar, Heilbron, Kim, Carey,
  Dreseler, and Li}{Alsubaiee et~al\mbox{.}}{2014b}]%
        {asterixdb-storage2014}
\bibfield{author}{\bibinfo{person}{Sattam Alsubaiee},
  \bibinfo{person}{Alexander Behm}, \bibinfo{person}{Vinayak Borkar},
  \bibinfo{person}{Zachary Heilbron}, \bibinfo{person}{Young-Seok Kim},
  \bibinfo{person}{Michael~J. Carey}, \bibinfo{person}{Markus Dreseler}, {and}
  \bibinfo{person}{Chen Li}.} \bibinfo{year}{2014}\natexlab{b}.
\newblock \showarticletitle{Storage Management in {AsterixDB}}.
\newblock \bibinfo{journal}{\emph{Proceedings of the VLDB Endowment (PVLDB)}}
  \bibinfo{volume}{7}, \bibinfo{number}{10} (\bibinfo{year}{2014}),
  \bibinfo{pages}{841--852}.
\newblock


\bibitem[\protect\citeauthoryear{Balmau, Didona, Guerraoui, Zwaenepoel, Yuan,
  Arora, Gupta, and Konka}{Balmau et~al\mbox{.}}{2017a}]%
        {triad2017}
\bibfield{author}{\bibinfo{person}{Oana Balmau}, \bibinfo{person}{Diego
  Didona}, \bibinfo{person}{Rachid Guerraoui}, \bibinfo{person}{Willy
  Zwaenepoel}, \bibinfo{person}{Huapeng Yuan}, \bibinfo{person}{Aashray Arora},
  \bibinfo{person}{Karan Gupta}, {and} \bibinfo{person}{Pavan Konka}.}
  \bibinfo{year}{2017}\natexlab{a}.
\newblock \showarticletitle{{TRIAD}: Creating Synergies Between Memory, Disk
  and Log in Log Structured Key-Value Stores}. In
  \bibinfo{booktitle}{\emph{USENIX Annual Technical Conference (ATC)}}.
  \bibinfo{pages}{363--375}.
\newblock


\bibitem[\protect\citeauthoryear{Balmau, Dinu, Zwaenepoel, Gupta,
  Chandhiramoorthi, and Didona}{Balmau et~al\mbox{.}}{2019}]%
        {silk2019}
\bibfield{author}{\bibinfo{person}{Oana Balmau}, \bibinfo{person}{Florin Dinu},
  \bibinfo{person}{Willy Zwaenepoel}, \bibinfo{person}{Karan Gupta},
  \bibinfo{person}{Ravishankar Chandhiramoorthi}, {and} \bibinfo{person}{Diego
  Didona}.} \bibinfo{year}{2019}\natexlab{}.
\newblock \showarticletitle{{SILK}: Preventing Latency Spikes in Log-Structured
  Merge Key-Value Stores}. In \bibinfo{booktitle}{\emph{USENIX Annual Technical
  Conference (ATC))}}. \bibinfo{pages}{753--766}.
\newblock


\bibitem[\protect\citeauthoryear{Balmau, Guerraoui, Trigonakis, and
  Zablotchi}{Balmau et~al\mbox{.}}{2017b}]%
        {flodb2017}
\bibfield{author}{\bibinfo{person}{Oana Balmau}, \bibinfo{person}{Rachid
  Guerraoui}, \bibinfo{person}{Vasileios Trigonakis}, {and}
  \bibinfo{person}{Igor Zablotchi}.} \bibinfo{year}{2017}\natexlab{b}.
\newblock \showarticletitle{{FloDB}: Unlocking Memory in Persistent Key-Value
  Stores}. In \bibinfo{booktitle}{\emph{European Conference on Computer Systems
  (EuroSys)}}. \bibinfo{pages}{80--94}.
\newblock


\bibitem[\protect\citeauthoryear{Bindschaedler, Goel, and
  Zwaenepoel}{Bindschaedler et~al\mbox{.}}{2020}]%
        {hailstorm2020}
\bibfield{author}{\bibinfo{person}{Laurent Bindschaedler},
  \bibinfo{person}{Ashvin Goel}, {and} \bibinfo{person}{Willy Zwaenepoel}.}
  \bibinfo{year}{2020}\natexlab{}.
\newblock \showarticletitle{Hailstorm: Disaggregated Compute and Storage for
  Distributed {LSM}-based Databases}. In
  \bibinfo{booktitle}{\emph{International Conference on Architectural Support
  for Programming Languages and Operating Systems (ASPLOS)}}.
  \bibinfo{pages}{301--316}.
\newblock


\bibitem[\protect\citeauthoryear{Bloom}{Bloom}{1970}]%
        {bloom-filter1970}
\bibfield{author}{\bibinfo{person}{Burton~H. Bloom}.}
  \bibinfo{year}{1970}\natexlab{}.
\newblock \showarticletitle{Space/Time Trade-offs in Hash Coding with Allowable
  Errors}.
\newblock \bibinfo{journal}{\emph{Communications of the ACM (CACM)}}
  \bibinfo{volume}{13}, \bibinfo{number}{7} (\bibinfo{date}{July}
  \bibinfo{year}{1970}), \bibinfo{pages}{422--426}.
\newblock


\bibitem[\protect\citeauthoryear{Bortnikov, Braginsky, Hillel, Keidar, and
  Sheffi}{Bortnikov et~al\mbox{.}}{2018}]%
        {accordion2018}
\bibfield{author}{\bibinfo{person}{Edward Bortnikov},
  \bibinfo{person}{Anastasia Braginsky}, \bibinfo{person}{Eshcar Hillel},
  \bibinfo{person}{Idit Keidar}, {and} \bibinfo{person}{Gali Sheffi}.}
  \bibinfo{year}{2018}\natexlab{}.
\newblock \showarticletitle{Accordion: Better Memory Organization for {LSM}
  Key-value Stores}.
\newblock \bibinfo{journal}{\emph{Proceedings of the VLDB Endowment (PVLDB)}}
  \bibinfo{volume}{11}, \bibinfo{number}{12} (\bibinfo{year}{2018}),
  \bibinfo{pages}{1863--1875}.
\newblock


\bibitem[\protect\citeauthoryear{Bu, Borkar, Xu, and Carey}{Bu
  et~al\mbox{.}}{2013}]%
        {bloat2013}
\bibfield{author}{\bibinfo{person}{Yingyi Bu}, \bibinfo{person}{Vinayak
  Borkar}, \bibinfo{person}{Guoqing Xu}, {and} \bibinfo{person}{Michael~J.
  Carey}.} \bibinfo{year}{2013}\natexlab{}.
\newblock \showarticletitle{A Bloat-Aware Design for Big Data Applications}. In
  \bibinfo{booktitle}{\emph{International Symposium on Memory Management
  (ISMM)}}. \bibinfo{pages}{119–130}.
\newblock


\bibitem[\protect\citeauthoryear{Cao, Liu, Wang, Chen, Zhu, Zheng, Wang, and
  Ma}{Cao et~al\mbox{.}}{2018}]%
        {polardb2018}
\bibfield{author}{\bibinfo{person}{Wei Cao}, \bibinfo{person}{Zhenjun Liu},
  \bibinfo{person}{Peng Wang}, \bibinfo{person}{Sen Chen},
  \bibinfo{person}{Caifeng Zhu}, \bibinfo{person}{Song Zheng},
  \bibinfo{person}{Yuhui Wang}, {and} \bibinfo{person}{Guoqing Ma}.}
  \bibinfo{year}{2018}\natexlab{}.
\newblock \showarticletitle{PolarFS: An Ultra-Low Latency and Failure Resilient
  Distributed File System for Shared Storage Cloud Database}.
\newblock \bibinfo{journal}{\emph{Proceedings of the VLDB Endowment (PVLDB)}}
  \bibinfo{volume}{11}, \bibinfo{number}{12} (\bibinfo{year}{2018}),
  \bibinfo{pages}{1849–1862}.
\newblock
\showISSN{2150-8097}


\bibitem[\protect\citeauthoryear{Carey}{Carey}{2019}]%
        {asterixdb2019}
\bibfield{author}{\bibinfo{person}{Michael~J. Carey}.}
  \bibinfo{year}{2019}\natexlab{}.
\newblock \showarticletitle{{AsterixDB} Mid-Flight: A Case Study in Building
  Systems in Academia}. In \bibinfo{booktitle}{\emph{International Conference
  on Data Engineering (ICDE)}}. \bibinfo{pages}{1--12}.
\newblock


\bibitem[\protect\citeauthoryear{Chang, Dean, Ghemawat, Hsieh, Wallach,
  Burrows, Chandra, Fikes, and Gruber}{Chang et~al\mbox{.}}{2008}]%
        {bigtable}
\bibfield{author}{\bibinfo{person}{Fay Chang}, \bibinfo{person}{Jeffrey Dean},
  \bibinfo{person}{Sanjay Ghemawat}, \bibinfo{person}{Wilson~C. Hsieh},
  \bibinfo{person}{Deborah~A. Wallach}, \bibinfo{person}{Mike Burrows},
  \bibinfo{person}{Tushar Chandra}, \bibinfo{person}{Andrew Fikes}, {and}
  \bibinfo{person}{Robert~E. Gruber}.} \bibinfo{year}{2008}\natexlab{}.
\newblock \showarticletitle{Bigtable: A distributed storage system for
  structured data}.
\newblock \bibinfo{journal}{\emph{ACM Transactions on Computer Systems}}
  \bibinfo{volume}{26}, \bibinfo{number}{2} (\bibinfo{year}{2008}),
  \bibinfo{pages}{4:1--4:26}.
\newblock


\bibitem[\protect\citeauthoryear{Chou and DeWitt}{Chou and DeWitt}{1986}]%
        {dbmin1986}
\bibfield{author}{\bibinfo{person}{Hong~Tai Chou} {and}
  \bibinfo{person}{David~J. DeWitt}.} \bibinfo{year}{1986}\natexlab{}.
\newblock \showarticletitle{An evaluation of buffer management strategies for
  relational database systems}.
\newblock \bibinfo{journal}{\emph{Algorithmica}} \bibinfo{volume}{1},
  \bibinfo{number}{1} (\bibinfo{date}{01 Nov} \bibinfo{year}{1986}),
  \bibinfo{pages}{311--336}.
\newblock
\showISSN{1432-0541}


\bibitem[\protect\citeauthoryear{Cooper, Silberstein, Tam, Ramakrishnan, and
  Sears}{Cooper et~al\mbox{.}}{2010}]%
        {ycsb2010}
\bibfield{author}{\bibinfo{person}{Brian~F. Cooper}, \bibinfo{person}{Adam
  Silberstein}, \bibinfo{person}{Erwin Tam}, \bibinfo{person}{Raghu
  Ramakrishnan}, {and} \bibinfo{person}{Russell Sears}.}
  \bibinfo{year}{2010}\natexlab{}.
\newblock \showarticletitle{Benchmarking Cloud Serving Systems with {YCSB}}. In
  \bibinfo{booktitle}{\emph{ACM Symposium on Cloud Computing (SoCC)}}.
  \bibinfo{pages}{143--154}.
\newblock


\bibitem[\protect\citeauthoryear{Dayan, Athanassoulis, and Idreos}{Dayan
  et~al\mbox{.}}{2017}]%
        {monkey2017}
\bibfield{author}{\bibinfo{person}{Niv Dayan}, \bibinfo{person}{Manos
  Athanassoulis}, {and} \bibinfo{person}{Stratos Idreos}.}
  \bibinfo{year}{2017}\natexlab{}.
\newblock \showarticletitle{Monkey: Optimal Navigable Key-Value Store}. In
  \bibinfo{booktitle}{\emph{ACM International Conference on Management of Data
  (SIGMOD)}}. \bibinfo{pages}{79--94}.
\newblock


\bibitem[\protect\citeauthoryear{Dayan, Athanassoulis, and Idreos}{Dayan
  et~al\mbox{.}}{2018}]%
        {monkey-tods}
\bibfield{author}{\bibinfo{person}{Niv Dayan}, \bibinfo{person}{Manos
  Athanassoulis}, {and} \bibinfo{person}{Stratos Idreos}.}
  \bibinfo{year}{2018}\natexlab{}.
\newblock \showarticletitle{Optimal {Bloom} Filters and Adaptive Merging for
  {LSM}-Trees}.
\newblock \bibinfo{journal}{\emph{ACM Transactions on Database Systems (TODS)}}
  \bibinfo{volume}{43}, \bibinfo{number}{4}, Article \bibinfo{articleno}{16}
  (\bibinfo{year}{2018}), \bibinfo{numpages}{16:1--16:48}~pages.
\newblock
\showISSN{0362-5915}


\bibitem[\protect\citeauthoryear{Dayan and Idreos}{Dayan and Idreos}{2018}]%
        {dostoevsky2018}
\bibfield{author}{\bibinfo{person}{Niv Dayan} {and} \bibinfo{person}{Stratos
  Idreos}.} \bibinfo{year}{2018}\natexlab{}.
\newblock \showarticletitle{Dostoevsky: Better Space-Time Trade-Offs for
  {LSM}-Tree Based Key-Value Stores via Adaptive Removal of Superfluous
  Merging}. In \bibinfo{booktitle}{\emph{ACM International Conference on
  Management of Data (SIGMOD)}}. \bibinfo{pages}{505--520}.
\newblock


\bibitem[\protect\citeauthoryear{Dayan and Idreos}{Dayan and Idreos}{2019}]%
        {lsm-bush}
\bibfield{author}{\bibinfo{person}{Niv Dayan} {and} \bibinfo{person}{Stratos
  Idreos}.} \bibinfo{year}{2019}\natexlab{}.
\newblock \showarticletitle{The Log-Structured Merge-Bush \& the Wacky
  Continuum}. In \bibinfo{booktitle}{\emph{ACM International Conference on
  Management of Data (SIGMOD)}}. \bibinfo{pages}{449--466}.
\newblock


\bibitem[\protect\citeauthoryear{Dong, Callaghan, Galanis, Borthakur, Savor,
  and Strum}{Dong et~al\mbox{.}}{2017}]%
        {rocksdb-space2017}
\bibfield{author}{\bibinfo{person}{Siying Dong}, \bibinfo{person}{Mark
  Callaghan}, \bibinfo{person}{Leonidas Galanis}, \bibinfo{person}{Dhruba
  Borthakur}, \bibinfo{person}{Tony Savor}, {and} \bibinfo{person}{Michael
  Strum}.} \bibinfo{year}{2017}\natexlab{}.
\newblock \showarticletitle{Optimizing Space Amplification in {RocksDB}}. In
  \bibinfo{booktitle}{\emph{Conference on Innovative Data Systems Research
  (CIDR)}}.
\newblock


\bibitem[\protect\citeauthoryear{Duan, Thummala, and Babu}{Duan
  et~al\mbox{.}}{2009}]%
        {ituned2009}
\bibfield{author}{\bibinfo{person}{Songyun Duan}, \bibinfo{person}{Vamsidhar
  Thummala}, {and} \bibinfo{person}{Shivnath Babu}.}
  \bibinfo{year}{2009}\natexlab{}.
\newblock \showarticletitle{Tuning Database Configuration Parameters with
  {iTuned}}.
\newblock \bibinfo{journal}{\emph{Proceedings of the VLDB Endowment (PVLDB)}}
  \bibinfo{volume}{2}, \bibinfo{number}{1} (\bibinfo{year}{2009}),
  \bibinfo{pages}{1246--1257}.
\newblock


\bibitem[\protect\citeauthoryear{Grover and Carey}{Grover and Carey}{2015}]%
        {asterixdb-feed2015}
\bibfield{author}{\bibinfo{person}{Raman Grover} {and}
  \bibinfo{person}{Michael~J. Carey}.} \bibinfo{year}{2015}\natexlab{}.
\newblock \showarticletitle{Data Ingestion in {AsterixDB}}. In
  \bibinfo{booktitle}{\emph{International Conference on Extending Database
  Technology (EDBT)}}. \bibinfo{pages}{605--616}.
\newblock


\bibitem[\protect\citeauthoryear{Huang, Cheng, Wang, Wang, He, Zhang, Li, Wang,
  Cao, and Li}{Huang et~al\mbox{.}}{2019}]%
        {xengine}
\bibfield{author}{\bibinfo{person}{Gui Huang}, \bibinfo{person}{Xuntao Cheng},
  \bibinfo{person}{Jianying Wang}, \bibinfo{person}{Yujie Wang},
  \bibinfo{person}{Dengcheng He}, \bibinfo{person}{Tieying Zhang},
  \bibinfo{person}{Feifei Li}, \bibinfo{person}{Sheng Wang},
  \bibinfo{person}{Wei Cao}, {and} \bibinfo{person}{Qiang Li}.}
  \bibinfo{year}{2019}\natexlab{}.
\newblock \showarticletitle{X-{Engine}: An Optimized Storage Engine for
  Large-scale {E-commerce} Transaction Processing}. In
  \bibinfo{booktitle}{\emph{ACM International Conference on Management of Data
  (SIGMOD)}}. \bibinfo{pages}{651--665}.
\newblock


\bibitem[\protect\citeauthoryear{Johnson and Shasha}{Johnson and
  Shasha}{1994}]%
        {2q}
\bibfield{author}{\bibinfo{person}{Theodore Johnson} {and}
  \bibinfo{person}{Dennis Shasha}.} \bibinfo{year}{1994}\natexlab{}.
\newblock \showarticletitle{{2Q}: A Low Overhead High Performance Buffer
  Management Replacement Algorithm}. In \bibinfo{booktitle}{\emph{International
  Conference on Very Large Data Bases (VLDB)}}. \bibinfo{pages}{439--450}.
\newblock


\bibitem[\protect\citeauthoryear{Kim, Behm, Blow, Borkar, Bu, Carey, Hubail,
  Jahangiri, Jia, Li, Luo, Maxon, and Pirzadeh}{Kim et~al\mbox{.}}{2020}]%
        {asterixdb-memory}
\bibfield{author}{\bibinfo{person}{Taewoo Kim}, \bibinfo{person}{Alexander
  Behm}, \bibinfo{person}{Michael Blow}, \bibinfo{person}{Vinayak Borkar},
  \bibinfo{person}{Yingyi Bu}, \bibinfo{person}{Michael~J. Carey},
  \bibinfo{person}{Murtadha Hubail}, \bibinfo{person}{Shiva Jahangiri},
  \bibinfo{person}{Jianfeng Jia}, \bibinfo{person}{Chen Li},
  \bibinfo{person}{Chen Luo}, \bibinfo{person}{Ian Maxon}, {and}
  \bibinfo{person}{Pouria Pirzadeh}.} \bibinfo{year}{2020}\natexlab{}.
\newblock \showarticletitle{Robust and efficient memory management in {Apache}
  {AsterixDB}}.
\newblock \bibinfo{journal}{\emph{Software: Practice and Experience}}
  \bibinfo{volume}{50}, \bibinfo{number}{7} (\bibinfo{year}{2020}),
  \bibinfo{pages}{1114--1151}.
\newblock


\bibitem[\protect\citeauthoryear{Li, Chan, Lee, and Xu}{Li
  et~al\mbox{.}}{2019a}]%
        {hashkv2019}
\bibfield{author}{\bibinfo{person}{Yongkun Li}, \bibinfo{person}{Helen H.~W.
  Chan}, \bibinfo{person}{Patrick P.~C. Lee}, {and} \bibinfo{person}{Yinlong
  Xu}.} \bibinfo{year}{2019}\natexlab{a}.
\newblock \showarticletitle{Enabling Efficient Updates in {KV} Storage via
  Hashing: Design and Performance Evaluation}.
\newblock \bibinfo{journal}{\emph{ACM Transactions on Storage (TOS)}}
  \bibinfo{volume}{15}, \bibinfo{number}{3}, Article \bibinfo{articleno}{20}
  (\bibinfo{year}{2019}).
\newblock


\bibitem[\protect\citeauthoryear{Li, Tian, Guo, Li, and Xu}{Li
  et~al\mbox{.}}{2019b}]%
        {elasticbf2019}
\bibfield{author}{\bibinfo{person}{Yongkun Li}, \bibinfo{person}{Chengjin
  Tian}, \bibinfo{person}{Fan Guo}, \bibinfo{person}{Cheng Li}, {and}
  \bibinfo{person}{Yinlong Xu}.} \bibinfo{year}{2019}\natexlab{b}.
\newblock \showarticletitle{{ElasticBF}: elastic bloom filter with hotness
  awareness for boosting read performance in large key-value stores}. In
  \bibinfo{booktitle}{\emph{{USENIX} Annual Technical Conference (ATC)}}.
  \bibinfo{pages}{739--752}.
\newblock


\bibitem[\protect\citeauthoryear{Lim, Andersen, and Kaminsky}{Lim
  et~al\mbox{.}}{2016}]%
        {lsm-model2016}
\bibfield{author}{\bibinfo{person}{Hyeontaek Lim}, \bibinfo{person}{David~G.
  Andersen}, {and} \bibinfo{person}{Michael Kaminsky}.}
  \bibinfo{year}{2016}\natexlab{}.
\newblock \showarticletitle{Towards Accurate and Fast Evaluation of Multi-Stage
  Log-structured Designs}. In \bibinfo{booktitle}{\emph{USENIX Conference on
  File and Storage Technologies (FAST)}}. \bibinfo{pages}{149--166}.
\newblock


\bibitem[\protect\citeauthoryear{Lu, Chen, Herodotou, and Babu}{Lu
  et~al\mbox{.}}{2019}]%
        {tuning2019}
\bibfield{author}{\bibinfo{person}{Jiaheng Lu}, \bibinfo{person}{Yuxing Chen},
  \bibinfo{person}{Herodotos Herodotou}, {and} \bibinfo{person}{Shivnath
  Babu}.} \bibinfo{year}{2019}\natexlab{}.
\newblock \showarticletitle{Speedup Your Analytics: Automatic Parameter Tuning
  for Databases and Big Data Systems}.
\newblock \bibinfo{journal}{\emph{Proceedings of the VLDB Endowment (PVLDB)}}
  \bibinfo{volume}{12}, \bibinfo{number}{12} (\bibinfo{year}{2019}),
  \bibinfo{pages}{1970--–1973}.
\newblock


\bibitem[\protect\citeauthoryear{Lu, Pillai, Arpaci-Dusseau, and
  Arpaci-Dusseau}{Lu et~al\mbox{.}}{2016}]%
        {wisckey2017}
\bibfield{author}{\bibinfo{person}{Lanyue Lu},
  \bibinfo{person}{Thanumalayan~Sankaranarayana Pillai},
  \bibinfo{person}{Andrea~C. Arpaci-Dusseau}, {and} \bibinfo{person}{Remzi~H.
  Arpaci-Dusseau}.} \bibinfo{year}{2016}\natexlab{}.
\newblock \showarticletitle{{WiscKey}: Separating Keys from Values in
  {SSD}-conscious Storage}. In \bibinfo{booktitle}{\emph{USENIX Conference on
  File and Storage Technologies (FAST)}}. \bibinfo{pages}{133--148}.
\newblock


\bibitem[\protect\citeauthoryear{Luo}{Luo}{2020}]%
        {lsm-memory2020}
\bibfield{author}{\bibinfo{person}{Chen Luo}.} \bibinfo{year}{2020}\natexlab{}.
\newblock \showarticletitle{Breaking Down Memory Walls in {LSM}-based Storage
  Systems}. In \bibinfo{booktitle}{\emph{ACM International Conference on
  Management of Data (SIGMOD)}}. \bibinfo{pages}{2817--2819}.
\newblock


\bibitem[\protect\citeauthoryear{Luo and Carey}{Luo and Carey}{2019a}]%
        {lsm-storage2019}
\bibfield{author}{\bibinfo{person}{Chen Luo} {and} \bibinfo{person}{Michael~J.
  Carey}.} \bibinfo{year}{2019}\natexlab{a}.
\newblock \showarticletitle{Efficient Data Ingestion and Query Processing for
  {LSM}-Based Storage Systems}.
\newblock \bibinfo{journal}{\emph{Proceedings of the VLDB Endowment (PVLDB)}}
  \bibinfo{volume}{12}, \bibinfo{number}{5} (\bibinfo{year}{2019}),
  \bibinfo{pages}{531--543}.
\newblock


\bibitem[\protect\citeauthoryear{Luo and Carey}{Luo and Carey}{2019b}]%
        {lsm-stability2019}
\bibfield{author}{\bibinfo{person}{Chen Luo} {and} \bibinfo{person}{Michael~J.
  Carey}.} \bibinfo{year}{2019}\natexlab{b}.
\newblock \showarticletitle{On Performance Stability in {LSM}-based Storage
  Systems}.
\newblock \bibinfo{journal}{\emph{Proceedings of the VLDB Endowment (PVLDB)}}
  \bibinfo{volume}{13}, \bibinfo{number}{4} (\bibinfo{year}{2019}),
  \bibinfo{pages}{449--462}.
\newblock


\bibitem[\protect\citeauthoryear{Luo and Carey}{Luo and Carey}{2020a}]%
        {lsm-memory-extended}
\bibfield{author}{\bibinfo{person}{Chen Luo} {and} \bibinfo{person}{Michael~J.
  Carey}.} \bibinfo{year}{2020}\natexlab{a}.
\newblock \showarticletitle{Breaking Down Memory Walls: Adaptive Memory
  Management in {LSM}-based Storage Systems (Extended Version)}.
\newblock \bibinfo{journal}{\emph{CoRR}}  \bibinfo{volume}{abs/2004.10360}
  (\bibinfo{year}{2020}).
\newblock


\bibitem[\protect\citeauthoryear{Luo and Carey}{Luo and Carey}{2020b}]%
        {lsm-survey}
\bibfield{author}{\bibinfo{person}{Chen Luo} {and} \bibinfo{person}{Michael~J.
  Carey}.} \bibinfo{year}{2020}\natexlab{b}.
\newblock \showarticletitle{{LSM}-based storage techniques: a survey}.
\newblock \bibinfo{journal}{\emph{The VLDB Journal (VLDBJ)}}
  \bibinfo{volume}{29}, \bibinfo{number}{1} (\bibinfo{year}{2020}),
  \bibinfo{pages}{393--418}.
\newblock


\bibitem[\protect\citeauthoryear{Luo, T{\"o}z{\"u}n, Tian, Barber, Raman, and
  Sidle}{Luo et~al\mbox{.}}{2019}]%
        {umzi2019}
\bibfield{author}{\bibinfo{person}{Chen Luo}, \bibinfo{person}{Pinar
  T{\"o}z{\"u}n}, \bibinfo{person}{Yuanyuan Tian}, \bibinfo{person}{Ronald
  Barber}, \bibinfo{person}{Vijayshankar Raman}, {and} \bibinfo{person}{Richard
  Sidle}.} \bibinfo{year}{2019}\natexlab{}.
\newblock \showarticletitle{Umzi: Unified Multi-Zone Indexing for Large-Scale
  {HTAP}}. In \bibinfo{booktitle}{\emph{International Conference on Extending
  Database Technology (EDBT)}}. \bibinfo{pages}{1--12}.
\newblock


\bibitem[\protect\citeauthoryear{Luo, Chatterjee, Ketsetsidis, Dayan, Qin, and
  Idreos}{Luo et~al\mbox{.}}{2020}]%
        {rosetta2020}
\bibfield{author}{\bibinfo{person}{Siqiang Luo}, \bibinfo{person}{Subarna
  Chatterjee}, \bibinfo{person}{Rafael Ketsetsidis}, \bibinfo{person}{Niv
  Dayan}, \bibinfo{person}{Wilson Qin}, {and} \bibinfo{person}{Stratos
  Idreos}.} \bibinfo{year}{2020}\natexlab{}.
\newblock \showarticletitle{{Rosetta}: A Robust Space-Time Optimized Range
  Filter for Key-Value Stores}. In \bibinfo{booktitle}{\emph{ACM International
  Conference on Management of Data (SIGMOD)}}. \bibinfo{pages}{2071–2086}.
\newblock


\bibitem[\protect\citeauthoryear{Mao, Jacobs, Amjad, Hristidis, Tsotras, and
  Young}{Mao et~al\mbox{.}}{2019}]%
        {lsm-append2019}
\bibfield{author}{\bibinfo{person}{Qizhong Mao}, \bibinfo{person}{Steven
  Jacobs}, \bibinfo{person}{Waleed Amjad}, \bibinfo{person}{Vagelis Hristidis},
  \bibinfo{person}{Vassilis~J Tsotras}, {and} \bibinfo{person}{Neal~E Young}.}
  \bibinfo{year}{2019}\natexlab{}.
\newblock \showarticletitle{Experimental Evaluation of Bounded-Depth {LSM}
  Merge Policies}. In \bibinfo{booktitle}{\emph{IEEE International Conference
  on Big Data}}. \bibinfo{pages}{523--532}.
\newblock


\bibitem[\protect\citeauthoryear{Mei, Cao, Jiang, and Li}{Mei
  et~al\mbox{.}}{2018}]%
        {sifrdb2018}
\bibfield{author}{\bibinfo{person}{Fei Mei}, \bibinfo{person}{Qiang Cao},
  \bibinfo{person}{Hong Jiang}, {and} \bibinfo{person}{Jingjun Li}.}
  \bibinfo{year}{2018}\natexlab{}.
\newblock \showarticletitle{{SifrDB}: A Unified Solution for Write-Optimized
  Key-Value Stores in Large Datacenter}. In \bibinfo{booktitle}{\emph{ACM
  Symposium on Cloud Computing (SoCC)}}. \bibinfo{pages}{477--489}.
\newblock


\bibitem[\protect\citeauthoryear{O'Neil, Cheng, Gawlick, and O'Neil}{O'Neil
  et~al\mbox{.}}{1996}]%
        {lsm1996}
\bibfield{author}{\bibinfo{person}{Patrick O'Neil}, \bibinfo{person}{Edward
  Cheng}, \bibinfo{person}{Dieter Gawlick}, {and} \bibinfo{person}{Elizabeth
  O'Neil}.} \bibinfo{year}{1996}\natexlab{}.
\newblock \showarticletitle{The Log-structured Merge-tree ({LSM}-tree)}.
\newblock \bibinfo{journal}{\emph{Acta Informatica}} \bibinfo{volume}{33},
  \bibinfo{number}{4} (\bibinfo{year}{1996}), \bibinfo{pages}{351--385}.
\newblock


\bibitem[\protect\citeauthoryear{O’Neil, O’Neil, and Weikum}{O’Neil
  et~al\mbox{.}}{1993}]%
        {lruk1993}
\bibfield{author}{\bibinfo{person}{Elizabeth~J. O’Neil},
  \bibinfo{person}{Patrick~E. O’Neil}, {and} \bibinfo{person}{Gerhard
  Weikum}.} \bibinfo{year}{1993}\natexlab{}.
\newblock \showarticletitle{The {LRU-K} Page Replacement Algorithm for Database
  Disk Buffering}.
\newblock \bibinfo{journal}{\emph{SIGMOD Record}} \bibinfo{volume}{22},
  \bibinfo{number}{2} (\bibinfo{date}{June} \bibinfo{year}{1993}),
  \bibinfo{pages}{297--306}.
\newblock


\bibitem[\protect\citeauthoryear{Qader, Cheng, and Hristidis}{Qader
  et~al\mbox{.}}{2018}]%
        {secondary2018}
\bibfield{author}{\bibinfo{person}{Mohiuddin~Abdul Qader},
  \bibinfo{person}{Shiwen Cheng}, {and} \bibinfo{person}{Vagelis Hristidis}.}
  \bibinfo{year}{2018}\natexlab{}.
\newblock \showarticletitle{A Comparative Study of Secondary Indexing
  Techniques in {LSM}-based {NoSQL} Databases}. In
  \bibinfo{booktitle}{\emph{ACM International Conference on Management of Data
  (SIGMOD)}}. \bibinfo{pages}{551--566}.
\newblock


\bibitem[\protect\citeauthoryear{Raju, Kadekodi, Chidambaram, and Abraham}{Raju
  et~al\mbox{.}}{2017}]%
        {pebblesdb2017}
\bibfield{author}{\bibinfo{person}{Pandian Raju}, \bibinfo{person}{Rohan
  Kadekodi}, \bibinfo{person}{Vijay Chidambaram}, {and} \bibinfo{person}{Ittai
  Abraham}.} \bibinfo{year}{2017}\natexlab{}.
\newblock \showarticletitle{{PebblesDB}: Building Key-Value Stores Using
  Fragmented Log-Structured Merge Trees}. In
  \bibinfo{booktitle}{\emph{Symposium on Operating Systems Principles (SOSP)}}.
  \bibinfo{pages}{497--514}.
\newblock


\bibitem[\protect\citeauthoryear{Ren, Zheng, Arulraj, and Gibson}{Ren
  et~al\mbox{.}}{2017}]%
        {slimdb2017}
\bibfield{author}{\bibinfo{person}{Kai Ren}, \bibinfo{person}{Qing Zheng},
  \bibinfo{person}{Joy Arulraj}, {and} \bibinfo{person}{Garth Gibson}.}
  \bibinfo{year}{2017}\natexlab{}.
\newblock \showarticletitle{{SlimDB}: A Space-efficient Key-value Storage
  Engine for Semi-sorted Data}.
\newblock \bibinfo{journal}{\emph{Proceedings of the VLDB Endowment (PVLDB)}}
  \bibinfo{volume}{10}, \bibinfo{number}{13} (\bibinfo{year}{2017}),
  \bibinfo{pages}{2037--2048}.
\newblock


\bibitem[\protect\citeauthoryear{Sacco and Schkolnick}{Sacco and
  Schkolnick}{1982}]%
        {hot-set1982}
\bibfield{author}{\bibinfo{person}{Giovanni~Maria Sacco} {and}
  \bibinfo{person}{Mario Schkolnick}.} \bibinfo{year}{1982}\natexlab{}.
\newblock \showarticletitle{A Mechanism for Managing the Buffer Pool in a
  Relational Database System Using the Hot Set Model}. In
  \bibinfo{booktitle}{\emph{International Conference on Very Large Data Bases
  (VLDB)}}. \bibinfo{pages}{257--262}.
\newblock


\bibitem[\protect\citeauthoryear{Sarkar, Papon, Staratzis, and
  Athanassoulis}{Sarkar et~al\mbox{.}}{2020}]%
        {lethe2020}
\bibfield{author}{\bibinfo{person}{Subhadeep Sarkar},
  \bibinfo{person}{Tarikul~Islam Papon}, \bibinfo{person}{Dimitris Staratzis},
  {and} \bibinfo{person}{Manos Athanassoulis}.}
  \bibinfo{year}{2020}\natexlab{}.
\newblock \showarticletitle{Lethe: A Tunable Delete-Aware {LSM} Engine}. In
  \bibinfo{booktitle}{\emph{ACM International Conference on Management of Data
  (SIGMOD)}}. \bibinfo{pages}{893–908}.
\newblock


\bibitem[\protect\citeauthoryear{Sears and Ramakrishnan}{Sears and
  Ramakrishnan}{2012}]%
        {blsm2012}
\bibfield{author}{\bibinfo{person}{Russell Sears} {and} \bibinfo{person}{Raghu
  Ramakrishnan}.} \bibinfo{year}{2012}\natexlab{}.
\newblock \showarticletitle{{bLSM}: A General Purpose Log Structured Merge
  Tree}. In \bibinfo{booktitle}{\emph{ACM International Conference on
  Management of Data (SIGMOD)}}. \bibinfo{pages}{217--228}.
\newblock


\bibitem[\protect\citeauthoryear{Storm, Garcia-Arellano, Lightstone, Diao, and
  Surendra}{Storm et~al\mbox{.}}{2006}]%
        {db2-memory2006}
\bibfield{author}{\bibinfo{person}{Adam~J. Storm}, \bibinfo{person}{Christian
  Garcia-Arellano}, \bibinfo{person}{Sam~S. Lightstone}, \bibinfo{person}{Yixin
  Diao}, {and} \bibinfo{person}{M. Surendra}.} \bibinfo{year}{2006}\natexlab{}.
\newblock \showarticletitle{Adaptive Self-tuning Memory in {DB2}}. In
  \bibinfo{booktitle}{\emph{International Conference on Very Large Data Bases
  (VLDB)}}. \bibinfo{pages}{1081--1092}.
\newblock


\bibitem[\protect\citeauthoryear{Tan, Zhang, Li, Chen, Zheng, Zhang, Qiao, Shi,
  Cao, and Zhang}{Tan et~al\mbox{.}}{2019}]%
        {ibtune2019}
\bibfield{author}{\bibinfo{person}{Jian Tan}, \bibinfo{person}{Tieying Zhang},
  \bibinfo{person}{Feifei Li}, \bibinfo{person}{Jie Chen},
  \bibinfo{person}{Qixing Zheng}, \bibinfo{person}{Ping Zhang},
  \bibinfo{person}{Honglin Qiao}, \bibinfo{person}{Yue Shi},
  \bibinfo{person}{Wei Cao}, {and} \bibinfo{person}{Rui Zhang}.}
  \bibinfo{year}{2019}\natexlab{}.
\newblock \showarticletitle{{IBTune}: Individualized Buffer Tuning for
  Large-Scale Cloud Databases}.
\newblock \bibinfo{journal}{\emph{Proceedings of the VLDB Endowment (PVLDB)}}
  \bibinfo{volume}{12}, \bibinfo{number}{10} (\bibinfo{year}{2019}),
  \bibinfo{pages}{1221–--1234}.
\newblock


\bibitem[\protect\citeauthoryear{Tran, Huynh, Tay, and Tung}{Tran
  et~al\mbox{.}}{2008}]%
        {tune-formula2008}
\bibfield{author}{\bibinfo{person}{Dinh~Nguyen Tran},
  \bibinfo{person}{Phung~Chinh Huynh}, \bibinfo{person}{Yong~C Tay}, {and}
  \bibinfo{person}{Anthony~KH Tung}.} \bibinfo{year}{2008}\natexlab{}.
\newblock \showarticletitle{A new approach to dynamic self-tuning of database
  buffers}.
\newblock \bibinfo{journal}{\emph{ACM Transactions on Storage (TOS)}}
  \bibinfo{volume}{4}, \bibinfo{number}{1} (\bibinfo{year}{2008}),
  \bibinfo{pages}{1--25}.
\newblock


\bibitem[\protect\citeauthoryear{Van~Aken, Pavlo, Gordon, and Zhang}{Van~Aken
  et~al\mbox{.}}{2017}]%
        {tune-ml2017}
\bibfield{author}{\bibinfo{person}{Dana Van~Aken}, \bibinfo{person}{Andrew
  Pavlo}, \bibinfo{person}{Geoffrey~J Gordon}, {and} \bibinfo{person}{Bohan
  Zhang}.} \bibinfo{year}{2017}\natexlab{}.
\newblock \showarticletitle{Automatic Database Management System Tuning Through
  Large-Scale Machine Learning}. In \bibinfo{booktitle}{\emph{ACM International
  Conference on Management of Data (SIGMOD)}}. \bibinfo{pages}{1009--1024}.
\newblock


\bibitem[\protect\citeauthoryear{Wang and Carey}{Wang and Carey}{2019}]%
        {idea2019}
\bibfield{author}{\bibinfo{person}{Xikui Wang} {and}
  \bibinfo{person}{Michael~J. Carey}.} \bibinfo{year}{2019}\natexlab{}.
\newblock \showarticletitle{An {IDEA}: An Ingestion Framework for Data
  Enrichment in {AsterixDB}}.
\newblock \bibinfo{journal}{\emph{Proceedings of the VLDB Endowment (PVLDB)}}
  \bibinfo{volume}{12}, \bibinfo{number}{11} (\bibinfo{year}{2019}),
  \bibinfo{pages}{1485--1498}.
\newblock


\bibitem[\protect\citeauthoryear{Yao}{Yao}{1978}]%
        {btree-space-1978}
\bibfield{author}{\bibinfo{person}{Andrew Chi-Chih Yao}.}
  \bibinfo{year}{1978}\natexlab{}.
\newblock \showarticletitle{On random 2--3 trees}.
\newblock \bibinfo{journal}{\emph{Acta Informatica}} \bibinfo{volume}{9},
  \bibinfo{number}{2} (\bibinfo{year}{1978}), \bibinfo{pages}{159--170}.
\newblock


\bibitem[\protect\citeauthoryear{Zhang, Liu, Zhou, Li, Xiao, Cheng, Xing, Wang,
  Cheng, Liu, et~al\mbox{.}}{Zhang et~al\mbox{.}}{2019}]%
        {tune-rl2019}
\bibfield{author}{\bibinfo{person}{Ji Zhang}, \bibinfo{person}{Yu Liu},
  \bibinfo{person}{Ke Zhou}, \bibinfo{person}{Guoliang Li},
  \bibinfo{person}{Zhili Xiao}, \bibinfo{person}{Bin Cheng},
  \bibinfo{person}{Jiashu Xing}, \bibinfo{person}{Yangtao Wang},
  \bibinfo{person}{Tianheng Cheng}, \bibinfo{person}{Li Liu}, {et~al\mbox{.}}}
  \bibinfo{year}{2019}\natexlab{}.
\newblock \showarticletitle{An End-to-End Automatic Cloud Database Tuning
  System Using Deep Reinforcement Learning}. In \bibinfo{booktitle}{\emph{ACM
  International Conference on Management of Data (SIGMOD)}}.
  \bibinfo{pages}{415--432}.
\newblock


\bibitem[\protect\citeauthoryear{Zhang, Wang, Cheng, Xu, Yu, Huang, Zhang, He,
  Li, Cao, Huang, and Sun}{Zhang et~al\mbox{.}}{2020}]%
        {fpga-lsm2020}
\bibfield{author}{\bibinfo{person}{Teng Zhang}, \bibinfo{person}{Jianying
  Wang}, \bibinfo{person}{Xuntao Cheng}, \bibinfo{person}{Hao Xu},
  \bibinfo{person}{Nanlong Yu}, \bibinfo{person}{Gui Huang},
  \bibinfo{person}{Tieying Zhang}, \bibinfo{person}{Dengcheng He},
  \bibinfo{person}{Feifei Li}, \bibinfo{person}{Wei Cao},
  \bibinfo{person}{Zhongdong Huang}, {and} \bibinfo{person}{Jianling Sun}.}
  \bibinfo{year}{2020}\natexlab{}.
\newblock \showarticletitle{{FPGA}-Accelerated Compactions for {LSM}-based
  Key-Value Store}. In \bibinfo{booktitle}{\emph{{USENIX} Conference on File
  and Storage Technologies (FAST)}}. \bibinfo{pages}{225--237}.
\newblock


\end{thebibliography}
	
\end{document}